\pdfoutput=1

\documentclass{iopart}

\usepackage{epsfig}
\usepackage{amscd,amssymb}
\usepackage{graphicx,xspace}
\usepackage{pstricks}
\usepackage[left=3cm,top=3cm,right=3cm,bottom=3cm]{geometry}
\usepackage{color}

\begin{document}


\title{Statistical Mechanics of Logarithmic REM: Duality, Freezing
and Extreme Value Statistics of $1/f$ Noises generated by Gaussian Free Fields}


\author{Yan V Fyodorov}

\address{School of Mathematical Sciences,
            University of Nottingham, Nottingham NG72RD, England}

\author{Pierre Le Doussal}

\address{CNRS-Laboratoire de Physique Th\'eorique de l'Ecole Normale Sup\'erieure\\
24 rue Lhomond, 75231 Paris
Cedex-France\thanks{LPTENS is a Unit\'e Propre du C.N.R.S.
associ\'ee \`a l'Ecole Normale Sup\'erieure et \`a l'Universit\'e Paris Sud}}

\author{Alberto Rosso }

\address{Laboratoire de Physique Th\'eorique et Mod\`eles Statistiques, CNRS (UMR 8626) \\
Universit\'e Paris-Sud, B\^at. 100, 91405 Orsay Cedex, France }

\date{Received:  / Accepted:  / Published }

\begin{abstract}
We compute the distribution of the partition functions for a
class of one-dimensional Random Energy Models (REM) with logarithmically
correlated random potential, above and at the glass transition
temperature. The random potential sequences represent various versions of the 1/f noise generated by sampling the two-dimensional Gaussian Free Field (2dGFF)
along various planar curves. Our method extends the recent analysis of \cite{FB}
from the circular case to an interval and is based on an
analytical continuation of the Selberg integral. In particular, we unveil
a {\it duality relation} satisfied by the suitable generating function of free energy cumulants in the high-temperature phase.
It reinforces the freezing scenario hypothesis for that generating function, from which we derive the distribution of
extrema for the 2dGFF on the $[0,1]$ interval. We
provide numerical checks of the circular and the interval case and
discuss universality and various extensions. Relevance to the distribution of
length of a segment in Liouville quantum gravity is noted.
\end{abstract}

\maketitle

\section{Introduction}

Describing the detailed statistics of the extrema of $M$ random
variables $V_i$ with logarithmic correlation built from those of
the two-dimensional Gaussian Free Field (2dGFF) $V(x)$, is a hard
and still mostly open problem. It arises in many fields from
physics and mathematics to finance.   The 2dGFF is a fundamental
object intimately related to conformal field theory
\cite{DiFrancesco}, and being also a building block of the
Liouville random measures $e^{V(z,\bar z)} dz d\bar z$ attracted
much interest in high-energy physics, quantum gravity, and pure
mathematics communities, see \cite{Qgrav} for an extensive list of
references.
 In the context of condensed matter physics the 2dGFF is of interest
to describe e.g. fluctuating interfaces between phases \cite{aarts}, e.g. their
confinement properties, multi-fractal properties of wave functions
of Dirac particle in random magnetic field \cite{chamon} and
associated Boltzmann-Gibbs measures \cite{Yan1}, glass transitions
of random energy models with logarithmic correlated energies
\cite{carpentier}, 2d self-gravitating systems \cite{selfgrav}
etc.. Descriptions of the level lines of the GFF as
Schramm-Loewner Evolutions (SLE) and conjectured relations to the
welding problem \cite{Jones} have also contributed in revival of
interest in the statistics of the GFF. In mathematical finance
there is a strong present interest in limit lognormal multifractal
processes \cite{bacry} (also called log-infinitely divisible
multifractal random measures), which is but a closely related
incarnation of the same object, see e.g. \cite{Vargas,Ostrov}.
Last, but not least important, is to look at the logarithmically correlated
random sequences as those representing various instances of $1/f$ noises, see  e.g. \cite{1fnoise} and \cite{FB}. Such noises regularly appear in many applications, and were recently discussed in the context of quantum chaos, where logarithmic correlations arise in sequences of energy levels \cite{qchaos}
or, as one can surmise, in the zeroes of the zeta Riemann function.
All this makes understanding extreme value statistics of such noises an interesting and important problem.

While the leading behavior $V_{min} \sim - 2 A \ln M$ is
rigorously proved \cite{mathGFF}, surprisingly little knowledge
exists on finer properties of the statistics of the GFF-related
minima, even heuristically. To serve this as well as many other
purposes it is of high interest to study the canonical partition
function $Z(\beta)=\sum_{i=1}^M e^{- \beta V_i}$ for the
corresponding Random Energy Model (REM) as a function of the
inverse temperature $\beta=1/T$. The distribution $P(F)$ of the
free energy $F = - T \ln Z$ reduces in the limit of zero
temperature $T=0$ to the distribution of the minimum $V_{min}$. A
few instances of REM can be solved explicitly, and are frequently
useful as approximations: (i) uncorrelated energies with variance
$\sim \ln M$, i.e. Derrida's original REM \cite{rem}, which gives
the correct constant $A$ \cite{chamon}(ii) paths with random
weights on trees, whose energies exhibit a similar logarithmic
scaling of correlations, but with a hierarchical structure rather
than a translationally invariant one \cite{derridaspohn,trees}
(iii) the infinite-dimensional Euclidean version of
logarithmically correlated REM and its further ramifications \cite
{Yan1,infdlog}. In particular, the close analogy of GFF-related
statistical mechanics with the models on trees
\cite{chamon,carpentier}, also noted in probability theory
\cite{mathGFF}, arises naturally in an approximate, i.e. one loop,
RG method, and led to the conjecture \cite{carpentier} that:
\begin{equation}
V_{\mbox{min}} = a_M+b_M y
\label{rescaledmin}
\end{equation}
with
\begin{eqnarray}
&& a_M= A (- 2 \ln M + \tilde \gamma \ln \ln M +O(1)) \quad, \quad b_M=A +O(1/\ln(M))
\label{am&bm}
\end{eqnarray}
where $\tilde \gamma=3/2$
and $y$ is a random variable of order unity whose probability density has
universal tails $p(y) \sim |y| e^y$ on the side $y \to -\infty$.
In addition it was convincingly demonstrated that the
log-correlated REM exhibits a freezing transition to a glass phase
dominated by a few minima, at the same $T_c$ as predicted by (i)
and (ii) \cite{carpentier,infdlog}. An outstanding problem left
fully open was to characterize the shape of the distribution of
the minimum beyond the tail, and in particular investigate whether
the universality also extends to that regime.

To address this issue, Fyodorov and Bouchaud \cite{FB} (FB) recently
considered a particular circular-log variant of REM.
Denoting here and henceforth the averaging over the random potential with the overbar,
the circular-log model is defined via the
correlation matrix $C_{ij}=\overline{V_i V_j}$ identical to those
of $M$ equidistant points $z_j=\exp(i \frac{2 \pi j}{M})$ on a
circle $C_{jk} = 2 G(z_j-z_k)$, where $G(z-z') = - \ln|z-z'|$ is
the full plane Green function of the 2dGFF. Equivalently, the above covariance function represents a $2\pi$-periodic real-valued Gaussian random process
$V(x)=\sum_{l=1}^{\infty}\left(v_l\,e^{ilx}+\bar{v}_l\,e^{-ilx}\right)$
with a {\it self-similar} spectrum $\langle v_l \bar{v}_m
\rangle=l^{-(2H+1)}\delta_{lm}$ characterised by the
particular choice of the Hurst exponent $H=0$.  Such a process
therefore represents a version of the so-called $1/f$ noise.

From the moments
$\overline{Z^n}$ FB reconstructed the distribution $P(Z)$ above
and at $T_c$. From such a point they proceeded by assuming that
for such a model the same freezing scenario as found in Ref.
\cite{carpentier} holds so that the generating function
\begin{eqnarray}\label{main}
&& g_\beta(y) = \overline{ \exp( - e^{\beta y} Z/Z_e  ) }, \quad
Z_e=M^{1+\beta^2}/\Gamma(1-\beta^2)
\end{eqnarray}
 remains in the thermodynamic limit $M\gg 1$ {\it
temperature independent} everywhere in the glass phase $T \leq
T_c$. As a result of such a conjecture they arrived at the
distribution of the minimum  of the random potential in their
problem. The corresponding probability density for the variable $y$ (defined in  (\ref{rescaledmin}) with $A=1$)
turned out to be given by
$p(y)=-g_{\infty}'(y)$ where
\begin{eqnarray}  \label{circ}
&& g_{\infty}(y)  = g_{\beta_c}(y)=2 e^{y/2}
K_1(2 e^{y/2})\,.
\end{eqnarray}
Such a density does indeed exhibit the universal Carpentier-Le Doussal tail
$p(y\to -\infty) \sim - y e^y$.

Our broad aim is to investigate analytically and numerically the
validity and universality of the above result, and to extend it to other models with logarithmic correlations.
In pursuing this goal we will be able, in particular, to
 extract statistics of the extrema of the (full plane) GFF sampled along an interval, $[0,1]$, with eventually some
charges at the endpoints of the interval. This breaks the circular symmetry of the
correlation matrix and one finds a different distribution. The
moments $\overline{Z^n}$ turn out to be given in some range of
positive integer $n$ by celebrated Selberg integrals
\cite{Selberg} \footnote{ In a somewhat different but related context this fact
was noticed, but not much exploited  in \cite{bacry}} and a first
(non-trivial) task is to analytically continue them to arbitrary
$n$. After suggesting a certain method for such a continuation
we are able to deduce the distribution of free energy $P(F)$ and
$g_{\beta}(y)$ at the freezing temperature $\beta=\beta_c$. The same conjecture as in FB
then yields the distribution of the minimum. As a by-product of our method we reveal a remarkable {\it duality property} enjoyed in the high-temperature phase
by the generating function precisely defined as in (\ref{main}) and unnoticed in \cite{FB}.
We conjecture such a duality
to be intimately related to the mechanisms behind freezing phenomenon. Finally we use direct
numerical simulations to verify the freezing scenario for the
circular ensemble and the resulting distribution (\ref{circ}), as
well as to test the new results of this paper for the interval case.
Universality and other cases are discussed at the end.


\section{Model and moments}

\subsection{Interval model}

Our starting point is the following continuum version of the
partition function of the Random Energy Model generated by a
Gaussian-distributed logarithmically-correlated random potential $V(x)$  defined on the interval $[0,1]$:
\begin{eqnarray} \label{1}
&& Z= \epsilon^{\beta^2} \int_{0}^{1} dx x^{a} (1-x)^b e^{- \beta V(x)}
\end{eqnarray}
with $a,b>-1$ real numbers and $\beta>0$.  The potential $V(x)$ is considered to have zero
mean and covariance inherited from the two-dimensional GFF:
\begin{eqnarray} \label{1a}
\overline{V(x) V(x')}=C(x-x')= - 2 \ln|x-x'|\,.
\end{eqnarray}
For the integral (\ref{1}) to be
well defined one needs to define a short scale cutoff $\epsilon \ll 1$. We therefore
tacitly assume in the expression (\ref{1a}) $V \to V_\epsilon$, with the
regularized potential being also Gaussian with a covariance function
 $C_\epsilon(x-x')$, such that the variance is $C_\epsilon(0)= 2 \ln(1/\epsilon)$.
We put for convenience the factor $\epsilon^{\beta^2}$ in front of the integral
to ensures that the integer moments $\overline{Z^n}$ are
$\epsilon-$independent in the high-temperature phase, see
Eq.(\ref{momentint}) below. At this stage we do not need to
specify the $\epsilon-$regularized form \footnote{there are
various useful cutoffs, e.g. the circle average, see e.g. \cite{Qgrav}, or the scale
invariant cone construction, see e.g. \cite{Vargas}}, but it is convenient for our purpose
below to require that $C_\epsilon(x)=C(x)$ for $|x|>\epsilon$.
Note that for $a=b=0$ the Gibbs measure of the disordered system
identifies with the random Liouville measure, and that $Z$ can be
interpreted as the (fluctuating) length of a segment in Liouville
quantum gravity see e.g. \cite{Qgrav} .

Below we will also consider a grid of $M$ points $x_i$, uniformly
spaced w.r.t the length element $dl=dx x^{a} (1-x)^b$ and the set
of values $V_i=V(x_i)$, $i=1,..M$. The correlation matrix
$\overline{V_i V_j}=C_{ij}$ at these grid values are $C_{ij}=-2
\ln(|i-j|/M)$ for $i \neq j$, and $C_{ii}=2 \ln M + W$ where
$W=\ln(1/(\epsilon M))$ is a constant of order unity, and we will
be interested in the limit \footnote{in practice we want that
$\min_i(|x_i-x_{i+1}|)>\epsilon$.} of large $M$ at fixed $\epsilon
M$. This generalizes the grid on the unit circle studied in
\cite{FB} where $x_j=e^{i \theta_j}$ with $\theta_j=2 \pi j/M$ and
$C_{ij}= C(x_i-x_j)= -2 \ln(2|\sin\frac{(\theta_i - \theta_j)}{2}|)$. We
will compare below the two situations. In each case one defines
the corresponding (discretized) REM by the partition function $Z_M=\sum_{i=1}^M e^{-\beta V_i}$.
We expect, as shown in \cite{FB} and discussed below, that there
is a sense in which universal features of the discretized version are
described by the continuum one in the large $M$ limit.

\subsection{positive moments}

Let us now compute the positive integer moments of $Z$. Denoting
$\gamma=\beta^2$, a straightforward calculation gives
\begin{eqnarray} \label{momentint}
&&  \overline{Z^n} =  \int_0^1\ldots \int_0^1 \prod_{i=1}^ndx_i
x_i^a (1-x_i)^b
  \prod _{1 \leq i < j \leq n} \frac{1}{|x_i-x_j|^{2 \gamma}}
\end{eqnarray}
where the small scale cutoff is implicit and modifies the
expressions for $|x_i-x_j|<\epsilon$. For a fixed $n=1,2,..$, a
well defined and universal $\epsilon\to 0$ limit exists whenever the
integral (\ref{momentint}) is convergent, in which case it is
given by the famous Selberg integral formula \cite{Selberg}
$\overline{Z^n}= s_n$, with:
\begin{eqnarray} \label{selberg0}
&& s_n(\gamma,a,b) = \prod_{j=1}^{j=n} \frac{\Gamma[1+a-(j-1)
\gamma] \Gamma[1+b-(j-1) \gamma] \Gamma(1-j
\gamma)}{\Gamma[2+a+b-(n+j-2) \gamma] \Gamma(1-\gamma)}
\end{eqnarray}
where $\Gamma(x)$ is the Euler gamma-function.
For $a,b>0$ the domain of convergence is given by $\gamma<1/n$. It
corresponds to the well known fact that for continuum REM models
 the distribution of $P(Z)$ develops algebraic
tails \footnote{for finite grid $1/M$ these tails are cut far
away by log-normal behaviour, see a detailed discussion in \cite{FB}}
hence integer moments $\overline{Z^n}$ become
infinite at a series of transition temperatures
$T_c^{(n)}=\sqrt{n}$. The true transition in the full Gibbs
measure happens however only at $T_c=1$ i.e. $\gamma=\gamma_c=1$.
Above $T_c$ the distribution $P(Z)$ exists in the limit
$\epsilon=0$, while the formally divergent moments start depending
on the cut-off parameter $\epsilon$. Analogous result arises in the
log-circular ensemble \cite{FB} where the moments of $Z_M$ were
analyzed, as recalled below. The generalizations for complex $a,b,\beta$, which
connect to sine-Gordon physics, as well as a detailed study of the competition with
binding transitions to the edges for $a,b<-1$ (in presence of a cutoff)
is mostly left for future studies, although some remarks about
the binding transitions are made below in Section \ref{sec:edge}
\footnote{the full conditions for convergence in (\ref{selberg0}) are $\Re(a), \Re(b)>-1$,
 $\Re(\gamma) < \min(1/n,(a+1)/(n-1), (b+1)/(n-1))$}.

\subsection{negative moments}
\label{neg}

Our first aim is to reconstruct the distribution $P(Z)$ from its
moments in the high temperature phase $\gamma \leq 1$. This
entails analytical continuation of the Selberg integral which is a
well known difficult problem. Here we present a solution of this
problem at $T_c$, the most interesting point. Let us first obtain
the negative integer moments for any $T \geq T_c$. It is
convenient to define:
\begin{equation}
z=\Gamma(1-\gamma) Z = e^{-\beta f}    \quad , \quad z_n = \overline{ z^n }
\end{equation}
which, as found below, and in \cite{FB}, has a well defined limit
as $T \to T_c^+$. One then checks for $a=b=0$ the following
recursion relation:
\begin{equation} \label{ratpos}
\frac{z_n}{z_{n-1}}=\frac{\Gamma[1-n\gamma]\,\Gamma^2[1-(n-1)
\gamma]\,\Gamma[2-(n-2)\gamma]}{\Gamma[2-(2n-3)\gamma]\,\Gamma[2-(2n-2)\gamma]}
\end{equation}
with $z_1=\Gamma(1-\gamma)$ (which also implies $z_0=1$), and a
similar formula for any $a,b$. Let us now perform the formal
analytic continuation to negative integer moments $m_k \equiv
z_{-k}$ in the above recursion (\ref{ratpos}) as $m_k/m_{k+1}
\equiv z_n/z_{n-1}|_{n\to -k}$. It is then easy to solve the
recursion starting from $m_0=z_0=1$. Restoring $a,b$ we find:
\begin{equation}\label{analcont}
z_{-k}= \prod_{j=1}^k \frac{\Gamma[2+a+b+(k+j+1)\gamma]}{
\Gamma[1+(j-1) \gamma]\,\Gamma[1+a+ j \gamma] \Gamma[1+b+ j \gamma]}
\end{equation}
We have checked that these expressions satisfy the convexity
property $z_n^{p-m} z_p^{m-n} \geq z_m^{p-n}$ for any integers
$n<m<p$ of arbitrary sign, which is a necessary condition for
positivity of a probability. For $a=b=0$ the formula
(\ref{analcont}) was announced very recently in \cite{Ostrov} as a
 rigorous consequence of certain recursion relations for Selberg
integrals.

Note that the domain in $a,b$ where (\ref{analcont}) remains well defined extends
to $a> -1-\gamma$, $b>-1-\gamma$, a region larger than the naive expectation
$a,b>-1$. This is a signature of the competition between binding to the edge and the random potential
as discussed below.

\subsection{From moments to distribution: the circular case and duality in the high-temperature phase}

Let us recall for comparison the corresponding analysis for the
circle\cite{FB}. There, the corresponding Dyson Coulomb gas
integrals give $z_n=\Gamma(1-n \gamma)$,  and such simple formula
admits the natural continuation to negative moments $n=-k$. This
allows to immediately and uniquely identify the distribution of
$1/z$ and leads to the probability densities:
\begin{equation} \label{gumbel}
P(z)= \beta^{-2} z^{- 1/\beta^2-1} \exp(-z^{-1/\beta^2}) \quad , \quad \tilde P(f)=\beta^{-1} \exp(f/\beta - e^{f/\beta})
\end{equation}
The latter formula implies that the free energy is distributed with a Gumbel probability
density for all $T \geq T_c$. Alternatively the (formal) series for positive
moments $g_\beta(y):=\overline{e^{-z e^{\beta y}}} =\sum_{n=0}^\infty \frac{(-1)^n}{n!} z_n
e^{n \beta y}$ is directly summed using $\Gamma(z)=\int_0^{\infty}e^{-t} t^{z-1}\,dt$ into the following generating function
\begin{equation}\label{12}
g_\beta(y) = \int_0^\infty dt\, \exp\{-t - e^{\beta y}\,
t^{-\beta^2}\}
\end{equation}
What went unnoticed in \cite{FB} was the remarkable duality relation satisfied by the exact expression for this function\footnote{In general such
duality holds for the transformation $\beta \to \beta_c^2/\beta$ but we specialized in this paper to $\beta_c=1$}:
\begin{equation}\label{circdual}
g_{\beta}(y)=g_{1/\beta}(y)
\end{equation}
To see this directly define $\tau=e^{\beta y}\,
t^{-\beta^2}$  implying $t=\tau^{-\frac{1}{\beta^2}}e^{-y/\beta}$, and after substituting this back to the integral
(\ref{12}) we see that
\begin{eqnarray}\label{12b}
 g_{\beta}(y) &=&-\frac{1}{\beta^2}
\int_0^\infty d\tau\, \tau^{-1-\frac{1}{\beta^2}}\,e^{ y/\beta}
\exp\{-\tau - e^{ y/\beta}\tau^{-\frac{1}{\beta^2}}\}\,\\ &=& \int_0^\infty d\tau \left[1+\frac{d}{d\tau}\right]
\exp\{-\tau - e^{ y/\beta}\tau^{-\frac{1}{\beta^2}}\} \equiv g_{\frac{1}{\beta}}(y)
\end{eqnarray}
as second term in the integrand  gives no contribution being full derivative of the expression vanishing at the boundaries of
the integration region. This transformation is formal in the sense that the function
$g_{1/\beta}(z)$ defined above for $\beta<1$ has nothing to do with the true generating function in the low temperature phase $\beta>1$.
 Rather, it is just obtained by taking the formula valid in the high temperature phase and changing everywhere $\beta \to 1/\beta$. However the duality
  relation still gives a precious information, e.g. it implies that an infinite set of derivatives $(\beta \partial_\beta)^n g_\beta(y)=0$ for any $n \geq 1$ odd at the self-dual point $\beta=1^-$. In particular the exact result:
\begin{eqnarray}
\partial_\beta g_\beta(y)|_{\beta=\beta_c^-} = 0 \quad , \quad {\rm for}~ {\rm all } ~~ y
\end{eqnarray}
shows that the "flow" of this function as a function of temperature vanishes at the critical point, quite consistent
with a freezing of the whole function (with continuous temperature derivatives). It is in fact quite amazing that precisely this generating function $g_\beta(y)= \overline{\exp(-e^{\beta y} z)}$, with precisely this built-in temperature dependence, is both conjectured to freeze and shown to be self-dual. It is thus tempting to conjecture that freezing and duality are related, i.e. it is $g_\beta(y)$ and no other variation of it (such as e.g. replacing $e^{\beta y}$ by any other function of both $y$ and $\beta$) which freezes {\it because} it is self-dual in the whole high temperature phase. The same type of self-duality relation, as we demonstrate below, extends to the interval case supporting the conjecture.

Unfortunately, the direct methods of resummation which work for the circular case fail for the more complicated problem at
hand, the interval $[0,1]$. For this reason one needs to develop a more general procedure, which is done below.

\subsection{From moments to distribution: generalities}

Instead here we now define the generic moments $M_{\beta}(s) =
\overline{z^{1-s}}$, $M_{\beta}(1)=1$ for any complex $s$, at
fixed inverse temperature $\beta$. In particular, the generating
function of the cumulants for the free energy $f=-\beta^{-1} \ln
z$ is related to $M_{\beta}(s)$ via
\begin{eqnarray}\label{cum1}
 \sum_{n=0}^\infty \frac{s^n}{n!} \beta^n \overline{f^n}^c = \ln M_{\beta}(1+s)
\end{eqnarray}
Definition of the probability density $P(z)$ implies the relation
\begin{eqnarray}
\int_{-\infty}^{+\infty} e^{2 t} P(e^t) e^{- s t}\, dt =
M_{\beta}(s)
\end{eqnarray}
which can be inverted as the contour integral:
\begin{eqnarray}\label{inv1}
 e^{- 2 t} P(e^{-t}) = \frac{1}{2 i \pi}\int  e^{- s t} M_{\beta}(s)\,ds \label{int1}
\end{eqnarray}
e.g. along a contour parallel to the imaginary axis $s=s_0+ i \omega$, provided the
integral is convergent, $s_0$ being chosen larger than any
singularity of the integrand.

Further using the definition (\ref{main}) the function
$g_{\beta}(y)$ is found to satisfy the identities
\begin{eqnarray}\label{rel2}
&& \beta \int_{-\infty}^{+\infty} e^{\beta y (s-1)} g_{\beta}(y)\, dy
= M_{\beta}(s) \Gamma(s-1)\\
&& g_{\beta}(y) = \beta^{-1} e^{\beta y} \frac{1}{2 i \pi}
\int e^{-s y} M_{\beta}(\frac{s}{\beta})
\Gamma(\frac{s}{\beta} -1)\, ds \label{int2}
\end{eqnarray}
Hence once we know $M_{\beta}(s)$ we can retrieve all the
interesting distributions. Moreover, relation (\ref{rel2}) defines
after integration by parts
 the generating function of the cumulants for the probability
density defined by $p_{\beta}(y)=-g_{\beta}'(y)$:
\begin{eqnarray}\label{cum2}
\sum_{n=1}^\infty \frac{s^n}{n!} \overline{y^n}^c \equiv \ln{\int_{-\infty}^{\infty} p_{\beta}(y)\,e^{ys}\, dy}= \ln
M_{\beta}(1+\frac{s}{\beta}) + \ln  \Gamma(1+\frac{s}{\beta})
\end{eqnarray}
Comparison with (\ref{cum1}) yields after recalling the series
expansion for $\ln  \Gamma(1+s)$ in terms of the Euler constant
$\gamma_E$ and Riemann zeta-function $\zeta(n)$ the following
model-independent relations:
\begin{equation}\label{cum}
\overline{y}=\overline{f}-\gamma_E T, \quad \overline{y^n}^c|_{n
\geq 2} =\overline{f^n}^c + (-1)^n (n-1)! \zeta(n) T^n\,,
\end{equation}
This relation is valid at all temperature and comes only from the
definition of $g_\beta(y)$. It is most useful at
$\beta=\beta_c=1$, if we accept the freezing scenario. Given that
in that case the l.h.s. freezes at its value at $\beta=1$ then we
easily retrieve all cumulants of the free energy for all $T \leq
T_c$ just from the knowledge of $g_{\beta=1}(y)$. Conversely, it
is useful to test the freezing hypothesis in numerics, as we will
see below.

 Let us now discuss how these moment relations reflect duality for the circular case.
 In the latter model $M_{\beta}(s)=\Gamma(1+ (s-1) \gamma)$,
hence from (\ref{cum2}) one finds:
\begin{eqnarray}\label{cum2a}
\sum_{n=1}^\infty \frac{s^n}{n!} \overline{y^n}^c =  \ln  \Gamma(1+ s \beta)
+ \ln  \Gamma(1+\frac{s}{\beta})
\end{eqnarray}
which is manifestly invariant by the formal transformation $\beta \to 1/\beta$.
The latter fact implies, via (\ref{cum2}), the self-duality for $p_{\beta}(y)$, hence for $g_{\beta}(y)$.
Such an indirect method of proving self-duality for  $g_{\beta}(y)$ has advantage when direct verification is
difficult in view of cumbersome and/or implicit form for the generating function in the whole high-temperature phase.
We shall see later on that it indeed works for the interval case.

\section{Analytical continuation at the critical temperature
and distribution of minima on the interval}

\subsection{no edge charges}

\begin{figure}[htpb]
  \centering
  \includegraphics[width=0.5\textwidth]{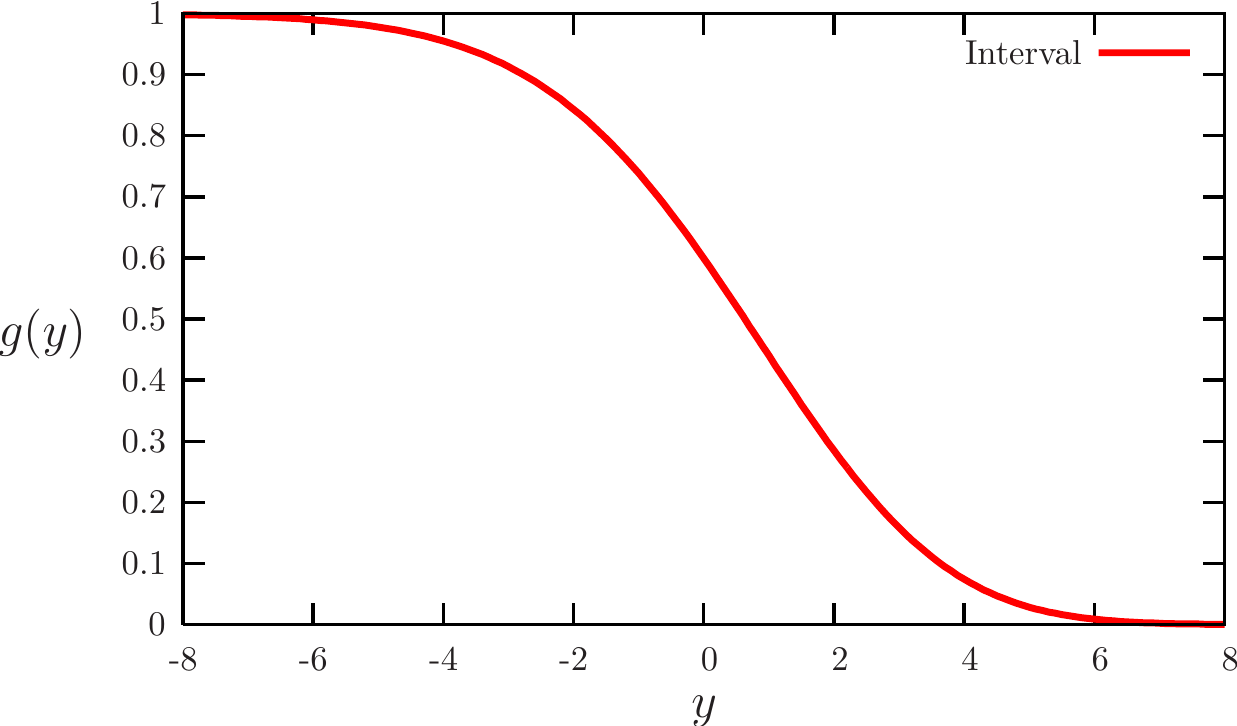}\includegraphics[width=0.5\textwidth]{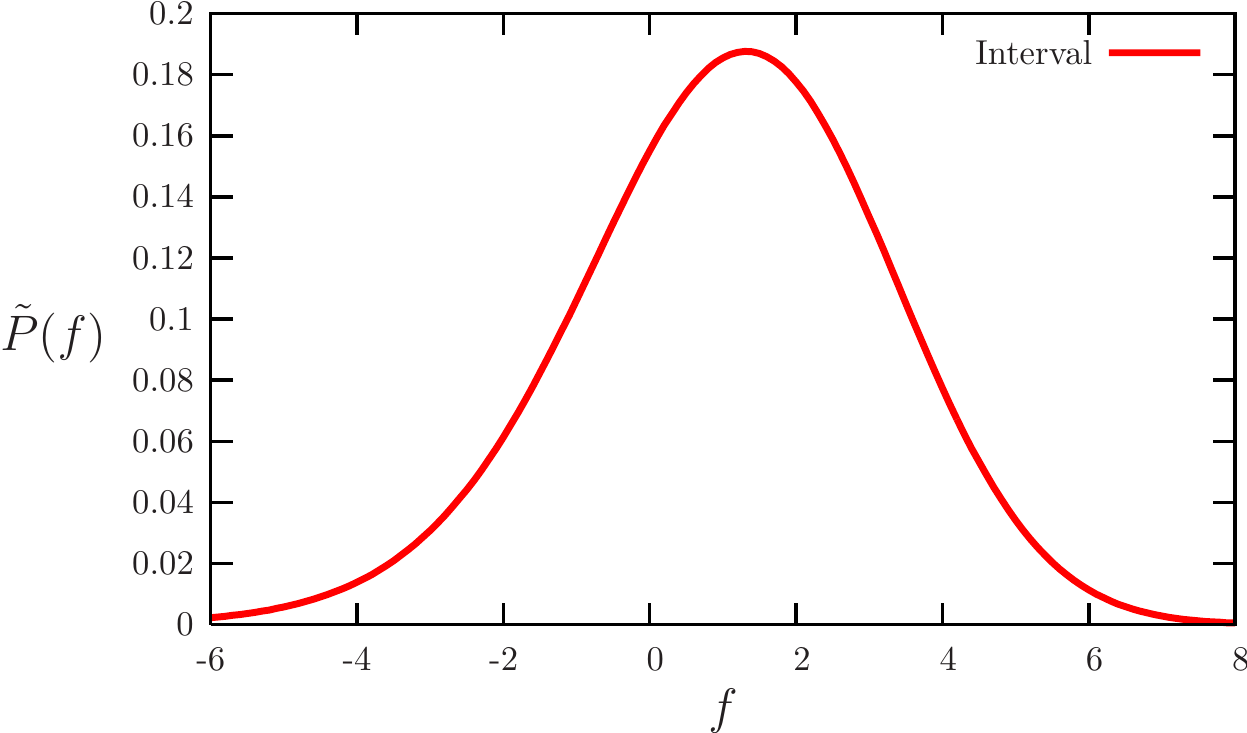}
  \caption{Color online. Analytical predictions for the interval $[0,1]$ with no edge charge:  (i) Left: plot of $g_{\beta_c}(y)$ which, according to the freezing scenario is also, up to a shift, the cumulative distribution of the minimum $V_{min}$
  (ii) Right:  the free energy density $\tilde P(f)$
  at the critical temperature $\beta_c$ for the interval. Both are obtained
  by the appropriate inverse Laplace transforms (\ref{inv1}) and (\ref{int2}) from the analytical continuation (\ref{res1}) of the moments as indicated in the text}
\label{fig:gfree}
\end{figure}

Let us keep focussing on the critical temperature $\beta=1$.
Denoting $M_{\beta=1}(s)\equiv M(s), g_{\beta=1}(y)\equiv g(y)$,
we start with $a=b=0$ case (no charges at the end of the interval)
 for the sake of simplicity. For negative integer values
$s=1-n$ one finds from (\ref{ratpos}) after exploiting the doubling
identity $\Gamma(2 z) = 2^{2 z-1} \Gamma(z)
\Gamma(1/2+z)/\sqrt{\pi}$ the relation:
\begin{eqnarray}
&&  \frac{M(s+1)}{M(s)} = 2^{3+ 4 s} (1+s)\frac{
\left[\Gamma(\frac{3}{2} + s)\right]^2 }{\pi \Gamma(s) \Gamma(3 +
s)} \label{rec01}
\end{eqnarray}
To continue this formula to any $s$ we will use the {\it Barnes function}, which under some mild conditions is the
only solution \cite{Barnes} of:
\begin{eqnarray} \label{defbarnes}
&& G(s+1) = G(s) \Gamma(s)
\end{eqnarray}
with $G(1)=1$. The Barnes function $G(s)$ is
meromorphic in the complex plane and has zeroes at all negative
integers $s=0, -1, -2, ...$ \cite{Barnes}. It can be computed as:
\begin{eqnarray}
&& G(z) = (2 \pi)^{(z-1)/2} e^{- \frac{1}{2} (z-1)(z-2) + \int_0^{z-1} dx x \psi(x) }
\end{eqnarray}
where $\psi(x)=\Gamma'(x)/\Gamma(x)$, the integral being on any contour not crossing the real negative axis.
Using (\ref{defbarnes}) one finds the following analytical continuation for the moments,
which is one of the main result of this paper:
\begin{equation} \label{res1}
\overline{z^{1-s}} = M(s) = \frac{2^{2 s^2 + s - 2} }{G(5/2)^2
\pi^{s-1}}  \frac{1}{\Gamma(s) \Gamma(s+2)}
\left[\frac{G(s+\frac{3}{2})}{G(s)}\right]^2
\end{equation}
with $G(5/2)=A^{-3/2} \pi^{3/4} e^{1/8} 2^{-23/24}$ where $A$ is
Glaisher-Kinkelin constant $A=e^{1/12 - \zeta'(-1)}=1.28242712$.
To guarantee that this is the correct continuation, we have
checked (i) positivity: $M(s)$ given above is finite and positive
on the interval $s \in [0,+\infty[$ i.e. all real moments
$n=1-s<1$ exist. (ii) convexity: on this interval $\partial_s^2
\ln M(s) >0$ (iii) convergence of the integrals
(\ref{int1},\ref{int2}) for $s_0>1$. The latter can be used to
compute $g_{\beta=1}(y)$ and $\tilde P(f)=e^{f}/(2 \pi i) \int e^{-f s} M(s) ds$,
which are plotted in Fig.\ref{fig:gfree}. Note
finally that it reproduces the negative integer moments (\ref{analcont}) for $a=b=0, \gamma=1$.

The free energy cumulants are to be determined from (\ref{cum2},\ref{cum}),
and one finds $<y>=\frac{7}{2}-2\gamma_E-\ln(2 \pi)$,
$<y^2>_c=\frac{4 \pi^2}{3}-\frac{27}{4}$, and for general $n \geq
3$:
\begin{eqnarray}
&& <y^n>_c = (-)^{n-1} (n-1)! \big( \zeta(n-1) (2^n-4) - \zeta(n) (2^n 3 -k) +2^{n+1}-1-2^{-n} \big)
\end{eqnarray}
with $k=4$ and the same formula for $<f^n>_c$ with $k=3$. As a
comparison for the circle $M(s)=\Gamma(s)$, hence $<y>=2 <f>= -2
\gamma_E$, and $<y^n>_c=2 <f^n>_c = 2 (-1)^n (n-1)! \zeta(n)$ for
$n \geq 2$.

An important property of $g(y)$ at criticality is its decay at $y
\to -\infty$. Deforming the integration contour in (\ref{int2})
one obtains $g(y)$ as a sum of residues over the (multiple) poles
of $M(s)$ at $s=-n$, which generates the expansion in powers of
$e^{y}$.
\begin{eqnarray} \label{res1a}
&& g(y) = 1 + (y + A') e^y + (A + B y +C y^2 + \frac{1}{6} y^3) e^{2 y} \\
&& +
e^y \sum_{n=2}^\infty \frac{1}{(2 n)!}  \partial_s^{2 n} e^{-s y} 2^{-(n+1)(2 n+3 + 4 s)} \pi^{n+1} \\ \nonumber &\times&
\frac{\Gamma(n+1+s)^{2 n+1} \Gamma(n+3+s) G(s+\frac{3}{2})^2 M(n+1+s)}{(s-1) s^2 (s+1)^4
..(s+n-1)^{2n-1} G(n+1+s+\frac{3}{2})^2} |_{s=-n} \nonumber
\end{eqnarray}
with $A'=2 \gamma_E + \ln(2 \pi)-1$ and $C = -0.253846$, $B =
1.25388$, $A = -5.09728$. Let us recall that for the circular
model the expression (\ref{circ}) implies:
\begin{eqnarray}
g^{(circ)}(y) = 1 + e^{y} ( y - 1 + 2 \gamma_E) + e^{2 y}
(\frac{1}{2} y - \frac{5}{4} +  \gamma_E) + ..
\end{eqnarray}
The behaviour $g(y)-1\sim y e^y$ is precisely the
universal tail found by Carpentier and Le Doussal \cite{carpentier}. It has its origin in the $1/z^2$ forward tail
which the probability density of $z$ develops at critical
$\beta=1$, with the first moment $<z>$ becoming infinite. On the
other side $y \to + \infty$ one expects much faster decay, for
example $g^{(circ)}(y) = \sqrt{\pi} e^{\frac{y}{4} -2 e^{y/2}} ( 1
+\frac{3}{16} e^{-y/2} +.. )$.

\subsection{extension to edge charges, binding transition}
\label{sec:edge}

Extending these considerations for any $a,b$, one finds:
\begin{eqnarray}
&& M(s)=2^{2 s^2 + s(1+ 2(a+b)) - 3 - 2 (a+b)}  \pi^{1-s}
 \frac{G(2+a) G(2+b) G(4+a+b)}{\Gamma(2+\frac{a+b}{2}) G(2+ \frac{a+b}{2})^2 G(\frac{5}{2}+ \frac{a+b}{2})^2} \nonumber  \\
&& \times \frac{\Gamma(1+\frac{a+b}{2} + s) G(1+ \frac{a+b}{2} +
s)^2 G(\frac{3}{2}+ \frac{a+b}{2} + s)^2}{ G(s) G(1+a+s) G(1+b+s)
G(3+a+b+s)} \label{result}
\end{eqnarray}
and checks again positivity and convexity for $s \in [0,+\infty[$ (for $a,b>-1$). We give only:
\begin{eqnarray} \label{cum2ab}
&& \overline{y^2}^c_{a,b} = \frac{\pi ^2}{6}+\gamma_E + 3 \phi(4 + a + b) - \phi(2+a) - \phi(2+b)
\end{eqnarray}
with $\phi(x) = \psi(x) + (x-1) \psi'(x)$, and the case $a=b$ in the limit $a \to + \infty$ where one then finds:
\begin{eqnarray} \label{cum23}
&& \overline{y^2}_{a,a}^c = \ln(8 a) +\frac{\pi ^2}{6}+\gamma +1 + O(a^{-1}) \nonumber \\
&& \overline{y^3}_{a,a}^c =  - \frac{\pi ^2}{3} - 2 \zeta(3)  + O(a^{-1}) +..
\end{eqnarray}
i.e. all cumulants have a limit except the second one. This limit is discussed again below.

A remarkable case is $a=b=-1/2$. Then a simplification occurs:
\begin{eqnarray}
&& M(s)=M_{-1/2,-1/2}(s) = 2^{2 s^2 - s - 1}  \pi^{1-s} \frac{\Gamma(\frac{1}{2} + s)}{s \Gamma(3/2)} \label{simpler}
\end{eqnarray}
One can trace this simplification to the fact that the structure of the correlation matrix
becomes much simpler in that case, as detailed in Appendix A.
The corresponding distribution is easlily found as (again this is for $\beta=\beta_c=1$):
\begin{eqnarray} \label{result0}
&& P(z) =  (\frac{\pi}{8})^{3/2} \frac{1}{\Gamma(3/2) z^2} \int_0^z  \frac{dz_1}{z_1^{3/2}} \int_{-\infty}^{+\infty} \frac{dt}{\sqrt{2 \pi}} e^{-\frac{t^2}{2} - 3 \sqrt{\ln 2} t - \frac{\pi}{8} \frac{1}{z_1} e^{-2 \sqrt{\ln 2} t}}
\end{eqnarray}
which reproduces the above moments, and behaves as $P(z) \sim \pi/z^2$ at large $z$. This yields, after some manipulations:
\begin{eqnarray} \label{result1}
g(y) =  \frac{\pi}{4}  \int_{-\infty}^{+\infty} \frac{dt}{\sqrt{2 \pi}} e^{-\frac{t^2}{2} - 2\sqrt{\ln 2} t}\int_{e^y}^{\infty}\left(1-\frac{e^y}{u}\right)e^{-\sqrt{\pi u/2}\, e^{-\sqrt{\ln 2} t}}\,du
\end{eqnarray}
and one finds $\overline{y^2}^c=4 \ln 2 - 3 + 2 \pi^2/3=6.35232$ and $\overline{y}=1-2 \gamma_E - \ln(\frac{\pi}{2})=-0.606014$ hence $\overline{f^2}^c=4 \ln 2 - 3 + \pi^2/2$.

Let us now discuss briefly the case $a,b<-1$, for simplicity we focus on $b=a$. In that case, the model
(\ref{1}) requires, at least naively, a short scale cutoff to avoid the divergence near the edges. However, from e.g. the discussion of Appendix D in \cite{carpentier} we know that there should be a competition between the random potential in the bulk and the binding effect by the edge: in presence of disorder it may be more favorable for the particle to explore the bulk and to remain unbound from the edge. It is quite nice that our analytical continuation captures that effect. As mentioned in Section \ref{neg}, from the negative moments one can guess that the complete domain over which the high temperature phase extends is:
\begin{eqnarray} \label{result1}
a \geq - 1 - \gamma \quad {\rm and} \quad \gamma \leq 1
\end{eqnarray}
with $\gamma=\beta^2$ (here $\beta_c=1$), where equality in the first condition corresponds to the binding transition to the edge, while the second
to the freezing transition. This implies in particular that for $\gamma=1$, the case studied above, the binding transition occurs at $a=-2$, and that for any larger value of $a$ the system should be at bulk critical freezing, with however some continuous dependence in $a$. We can indeed check that the result (\ref{result}) for $M(s)$ leads to a well defined probability $P(z)$ for any $a>-2$. For instance one sees that the formula (\ref{cum2ab}) yields a finite $\overline{y^2}^c$ for any $a>-2$, which however diverges as $a \to -2^+$. The domain of definition becomes $s> -1-a$ for $-1>a>-2$, as the resulting $P(z)$ acquires now a broader tail $\sim 1/z^{3+a}$ at large $z$, while it was $\sim 1/z^2$ for $a>-1$. As $a \to -2$ the tail becomes non-normalizable as $\sim 1/z$, a signature of the binding transition. The case $a=-3/2$ provides a good illustration as (\ref{result}) again simplifies into:
\begin{eqnarray}
&& M(s) = M_{-3/2,-3/2}(s)=2^{2 s^2 - 5 s + 3}  \pi^{1/2-s} \Gamma(s-\frac{1}{2}) \label{simpler2}
\end{eqnarray}
which implies that the random variable $z$ can be written $z=z_1 e^{-f_2}$ where $z_1>0$ and $f_2$ are two independent random variables, $f_2$ being gaussian distributed with $\overline{f_2}=- \ln(2 \pi)$ and $\overline{f_2^2}^c=4 \ln 2$, and $z_1$ with distribution $P_1(z_1)=z_1^{-3/2} e^{-1/z_1}/\sqrt{\pi}$, leading to
the explicit form:
\begin{eqnarray}
P(z) = \frac{1}{z^{3/2} \pi \sqrt{8 \ln 2}} \int_{-\infty}^\infty dt \exp(-\frac{3}{2} t - \frac{1}{z} e^{-t} - \frac{(t+\ln(2 \pi))^2}{8 \ln2})
\end{eqnarray}
which does exhibit the $\sim 1/z^{3/2}$ tail at large $z$. We leave further studies of the
global phase diagram for arbitrary $a,b$ to the future.

\section{High temperature phase for $[0,1]$ interval with no end charges}

Let us consider the segment $[0,1]$ at any $\beta \leq \beta_c=1$, i.e. $\gamma=\beta^2<1$.
The moments must satisfy (using again the doubling identity):
\begin{eqnarray} \label{recT}
&& \frac{M_{\beta}(s+1)}{M_{\beta}(s)} = \frac{2^{2+\gamma+4 s \gamma}}{\pi} \frac{\Gamma(\frac{3}{2} + s \gamma) \Gamma(1+ \frac{\gamma}{2} + s \gamma) \Gamma(\frac{3}{2} + \frac{\gamma}{2} + s \gamma)}{ \Gamma(1-\gamma + s \gamma) \Gamma(2+\gamma + s \gamma) \Gamma(1+s \gamma)}
\end{eqnarray}
We need to find a way of continuing the moments to the complex plane.
To this end we define the function $G_\beta(x)$ for $\Re(x)>0$ by \cite{Zamo}:
\begin{eqnarray}\label{ZamoBarn}
&&  \! \! \! \! \! \! \! \! \! \! \! \! \! \! \! \!  \! \! \! \! \! \! \! \! \! \! \! \! \! \! \! \! \ln G_\beta(x) = \frac{x-Q/2}{2} \ln (2 \pi) + \int_0^\infty \frac{dt}{t} \big( \frac{e^{- \frac{Q}{2} t} - e^{- x t}}{(1-e^{-\beta t})(1-e^{-t/\beta})}
+\frac{e^{-t}}{2} (Q/2-x)^2 + \frac{Q/2-x}{t} \big)
\end{eqnarray}
where $Q=\beta+1/\beta$. This function is self-dual:
\begin{eqnarray}\label{ZamoBarndual}
&& G_\beta(x) = G_{1/\beta}(x)
\end{eqnarray}
and satisfies the property that we need, see e.g. \cite{Zamo}
and Appendix B,
\begin{eqnarray} \label{Gt1}
&&G_\beta(x + \beta) = \beta^{1/2 - \beta x}(2 \pi)^{\frac{\beta-1}{2}} \Gamma(\beta x)\,G_\beta(x)
\end{eqnarray}
One can check that $G_\beta(x)$ for $\beta=\beta_c=1$ coincides with the Barnes function $G(x)$ defined in the previous Section, e.g. setting $\beta=1$ in (\ref{Gt1}) one sees that $G_1(x + 1)=\Gamma(x)G_1(x)$, and, using $Q=2$ we have $G_1(1)=1$. Similarly to the standard Barnes function the new function $G_\beta(x)$ has no poles and only zeroes, and these are located at $x=-n \beta  - m/\beta$, $n,m=0,1,..$. It provides us with a natural generalization which can be used to perform the required analytical continuation for any temperature.

Using the above properties we find that
\begin{eqnarray} \label{MT}
&&  \! \! \! \! \! \! \! \! \! \! \! \! \! \! \! \! \! \! \! \! \! \! \! \! \! \! \! \! \! \! \! \!   M_{\beta}(s) = A_\beta
2^{(s-1)(2+\beta^2 (2 s+1))} \pi^{1-s} \frac{ \Gamma(1+ \beta^2 (s-1)) G_\beta(\frac{\beta}{2} + \frac{1}{\beta} + \beta s)
G_\beta(\frac{3}{2 \beta} + \beta s) G_\beta(\frac{\beta}{2} + \frac{3}{2 \beta} + \beta s) }{G_\beta( \beta + \frac{2}{\beta} + \beta s)
G^2_\beta(\frac{1}{\beta} + \beta s)} \nonumber \\
&&
\end{eqnarray}
with
\begin{eqnarray}
&& A_\beta=\frac{G_\beta(\frac{1}{\beta} + \beta)^2 G_\beta( 2 \beta + \frac{2}{\beta})}{G_\beta(\frac{3 \beta}{2} + \frac{1}{\beta}) G_\beta(\frac{3}{2 \beta} + \beta) G_\beta(\frac{3 \beta}{2} + \frac{3}{2 \beta}))}
\end{eqnarray}
reproduces correctly the recursion relation (\ref{recT}), hence provides an analytical continuation for the moments valid for $\beta<\beta_c=1$. We have checked numerically that it does satisfy positivity, convexity and a convergent inverse Laplace transform from which one can compute $P(z)$ and $g_\beta(y)$ using (\ref{int1}) (\ref{int2}). We will not study these
in details here, but give only a few properties.

Let us first check the duality. One easily sees that if one defines
\begin{eqnarray}
&& M_\beta(s) = 2^{1-s} \tilde M_\beta(s)
\end{eqnarray}
then $\ln  \tilde M_\beta(1+ \frac{s}{\beta}) + \ln  \Gamma(1+\frac{s}{\beta})$ is fully invariant
under $\beta \to 1/\beta$. From (\ref{cum2}) it implies that all $\overline{y^n}^c$ with $n \geq 2$ are
invariant by duality, only the average $\overline{y}$ is not. This is not a problem since this
average is not expected to be universal, and is easily remedied by defining
$\tilde z=z/2$ (which could have been done from the start)
and $\tilde g_\beta(y) = \overline{ \exp(- e^{\beta y}z/2) }$. Hence we
conclude that up to such a trivial shift the probability $\tilde p_\beta(y)=-\tilde g'_\beta(y)$ is
self dual, i.e. $\tilde p_{1/\beta}(y) = \tilde p_\beta(y)$.  From the discussion in the previous Section we conjecture that it is this function which freezes at $\beta=\beta_c=1$.

From the result (\ref{MT}) we can extract the cumulants of the free energy using (\ref{cum1}). We only discuss here the lowest non-trivial cumulant, given by
\begin{eqnarray} \label{f2int}
&& \overline{f^2}^c =  \overline{y^2}^c - \frac{\pi^2}{6} T^2 =  \frac{1}{\beta^2} \partial^2_s \ln M_{\beta}(1+s)|_{s=0} \\
&& \! \! \! \! \! \! \! \! \! \! \! \! \! \! \! \! = 4 \ln 2 +  h_\beta(\frac{3 \beta}{2} + \frac{1}{\beta})  + h_\beta(\beta + \frac{3}{2 \beta})
+ h_\beta(\frac{3 \beta}{2} + \frac{3}{2 \beta}) -
2 h_\beta(\beta + \frac{1}{\beta}) - h_\beta(2 \beta + \frac{2}{\beta}) + \frac{\beta^2 \pi^2}{6} \nonumber \\
\end{eqnarray}
where we have defined the self-dual function (see Appendix B):
\begin{eqnarray}\label{h1b}
\! \! \! \! \! \! \! \! \! \! \! \! \! \! \! \!  h_\beta(x)= h_{1/\beta}(x) = \partial_x^2 \ln G_\beta(x) = \ln{x}+ \int_0^\infty \frac{dt}{t}  e^{-xt} \left(1-\frac{t^2}{(1-e^{-\beta t})(1-e^{-t/\beta})} \right)
\end{eqnarray}
and we have used $\psi'(1)=\pi^2/6$. The resulting curve $\overline{f^2}^c$ as a function of
$\beta$ is plotted in Fig. \ref{fig:f2}. One finds that it increases from $\overline{f^2}^c(\beta \to 0)=3$
to $\overline{f^2}^c(\beta=1) = 7 \pi^2/6 - 27/4 = 4.76454$. More discussion is given in Section
and Appendix C, together with high temperature expansions.

\section{Gaussian weight model}

We now briefly discuss a case where the above considerations fail, and present
below some hints of why this may happen.

We consider now the continuum partition function for the log-correlated field on
the full real axis but with a gaussian weight:
$$Z=\epsilon^{\beta^2} \frac{1}{\sqrt{2 \pi}} \int_{-\infty}^\infty dx ~  e^{-x^2/2} e^{- \beta V(x)}$$
This problem is appealing as it leads to Mehta integrals and moments $z^{(G)}_n= \overline{z^n}= \overline{Z^n \Gamma(1-\beta^2)^n} = \prod_{j=1}^{j=n}
\Gamma[1- j \beta^2]$, i.e. simpler expressions than for the interval case considered above.

At criticality $\beta=1$ this implies
$M^{(G)}(s+1)/M^{(G)}(s)=1/\Gamma(s)$ for $s=-n$, which naturally
suggests $M^{(G)}(s)=1/G(s)$. This is positive for $s>0$ but,
surprisingly, convexity fails for for $s>s_c=1.92586..$. Hence this is not
an acceptable analytic continuation.

To get another handle on the problem one notes that this model can be obtained
from the large $a$ limit of the interval problem $[0,1]_{aa}$. Writing $x=1/2+y$
and performing the change of variable in (\ref{momentint}) one finds:
\begin{equation}
\lim_{a \to + \infty} (2 \pi)^{- n/2} 2^{2 a n} (8 a)^{\frac{n}{2}
- \frac{n(n-1)}{2} \gamma} z_n(a,a,\gamma) = z^{(G)}_n(\gamma)
\end{equation}
Not surprisingly one finds that the pointwise limit:
\begin{equation}
M^{(G)}(s) = \lim_{a \to + \infty} (2 \pi)^{- (1-s)/2} 2^{2 a
(1-s)} (8 a)^{\frac{1}{2}(1-s^2)}  M_{a,a}(s)
\end{equation}
yields $1/G(s)$ as expected. From this we also get that for large $a$:
\begin{equation}
\partial_s^2 \ln M^{(G)}(s) = - \ln(8 a) + \partial_s^2 \ln M_{a,a}(s)
\end{equation}
While the second term is nicely positive for all $s>0$ the
additional factor $- \ln(8 a)$ makes the total sum negative for
$s>s_c$, violating convexity. In other words while $M_{a,a}(s)$ corresponds to a well
defined distribution of probability, corresponding to the problem on the interval with edge charges, $M_G(s)$ corresponds then
to this probability "convoluted by a gaussian of negative
variance" and fails to be a probability. Note that such a shift in the second cumulant
$\overline{y^2}$ is indeed needed to obtain a finite final result in
(\ref{cum23}). All higher cumulants $\overline{y^n}^c|_{aa}$ with
$n \geq 3$ have a nice finite limit as $a \to \infty$, and can be extracted
from the generating function
\begin{eqnarray}
&& \sum_{n=0}^\infty \frac{s^n}{n!} \overline{y^{1+n}}^c =  \frac{1}{s} - (s-1) \psi(s) + s -
\frac{1}{2} \ln(2 \pi) - \frac{1}{2} \nonumber
\end{eqnarray}
obtained from $1/G(s)$. Hence the main problem seems to lie in the
second cumulant, and one may speculate that it is related to an inadequate
treatment of zero mode fluctuations. Another (possibly related) observation is that for $a>>1$ the whole
contribution to the $[0,1]_{aa}$ integral comes from a very small
vicinity (of the widths of $L_a\sim 1/\sqrt{a}$ ) of the mid-point
$x=1/2$ of the integration domain. One expects a competition between $L_a$ and the
regularization scale for the logarithm, so it may be that the result depends
on the order of limits $\epsilon\to 0$ and $a\to \infty$. We leave further study of this
problem to the future and now turn to numerical studies.

\section{Numerical study}

\subsection{circular ensemble}

We now turn to the numerical checks for the random
variables $V_i$ on $i=1,..M$ grid points and their associated REM of partition function $Z_M=\sum_{i=1}^M e^{-\beta V_i}$. We start with the log-circular ensemble
and study the $M \times M$ cyclic correlation matrix (choosing here $W=0$):
\begin{equation} \label{circmat}
C_{ij} =  -2 \ln(2|\sin \frac{ \pi (i-j)}{M} |) \quad i \neq j,  \quad C_{ii} = 2 \ln M + W
\end{equation}
whose eigenvalues $\lambda_k=2\ln{M}-2\sum_{n=1}^{M-1}\cos\{\frac{2\pi}{M}nk\}
\ln\{2\sin{\frac{\pi}{M}n}\}$ are all positive, with the uniform mode $\lambda_0=0$ for any $M$.
Let us recall that the relation to the continuum model defined above was established in \cite{FB} where it was shown that at large $M$ one has $\overline{Z_M^n} = z_n \overline{Z_e^n}$ for $\beta^2 n<1$ and $\overline{Z_M^n} \sim M^{1+n^2 \beta^2}$ for $\beta^2 n>1$ (the positive moments which formally diverge in the continuum).

The random variables $V_i$ are generated (for $M$ even) as $$V_l=\sqrt{\frac{2}{M}}\sum_{k=1}^{M/2}\sqrt{\lambda _k}\left[x_k\cos\{\frac{2\pi}{M}kl\}+y_k\sin\{\frac{2\pi}{M}kl\}\right]$$ where the $x_k$ and $y_k$ are two uncorrelated sets of i.i.d. real unit centered Gaussian variables. This is done using Fast Fourier Transform (FFT).

\begin{figure}[htpb]
  \centering
  \includegraphics[width=0.45\textwidth]{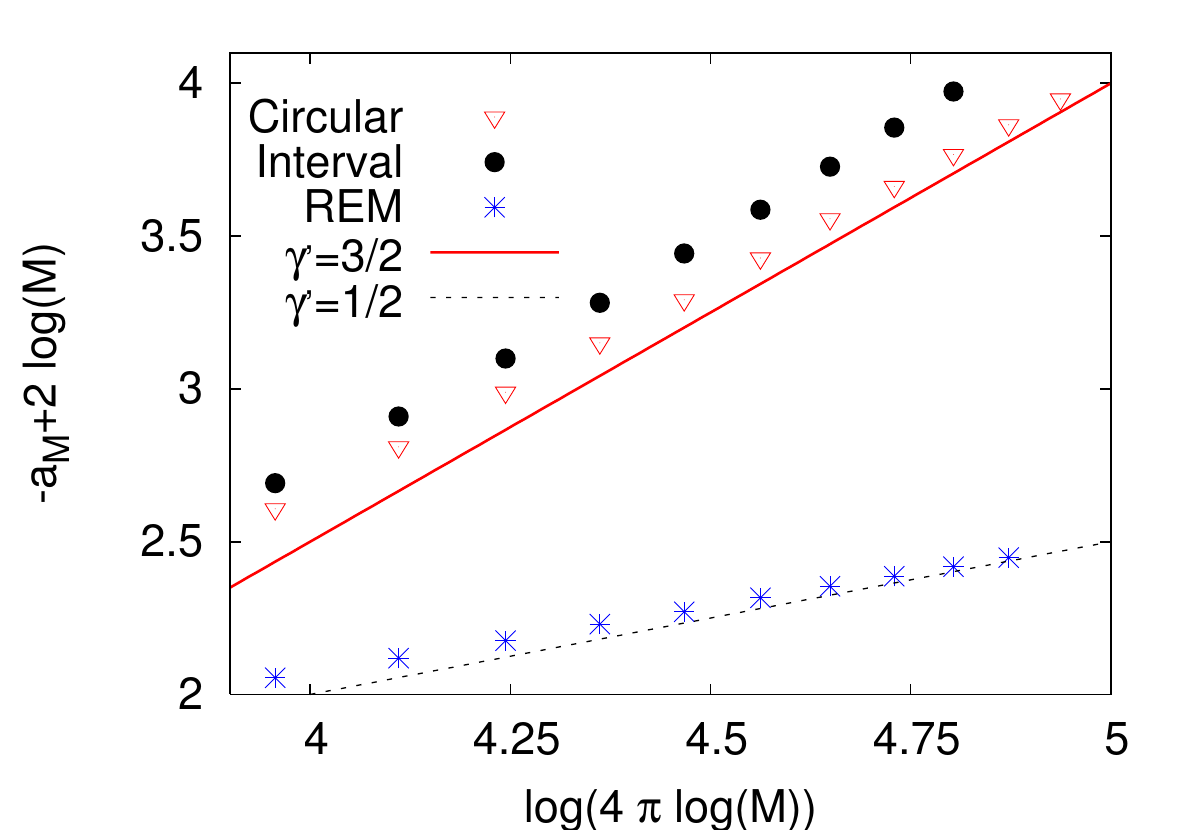} \includegraphics[width=0.45\textwidth]{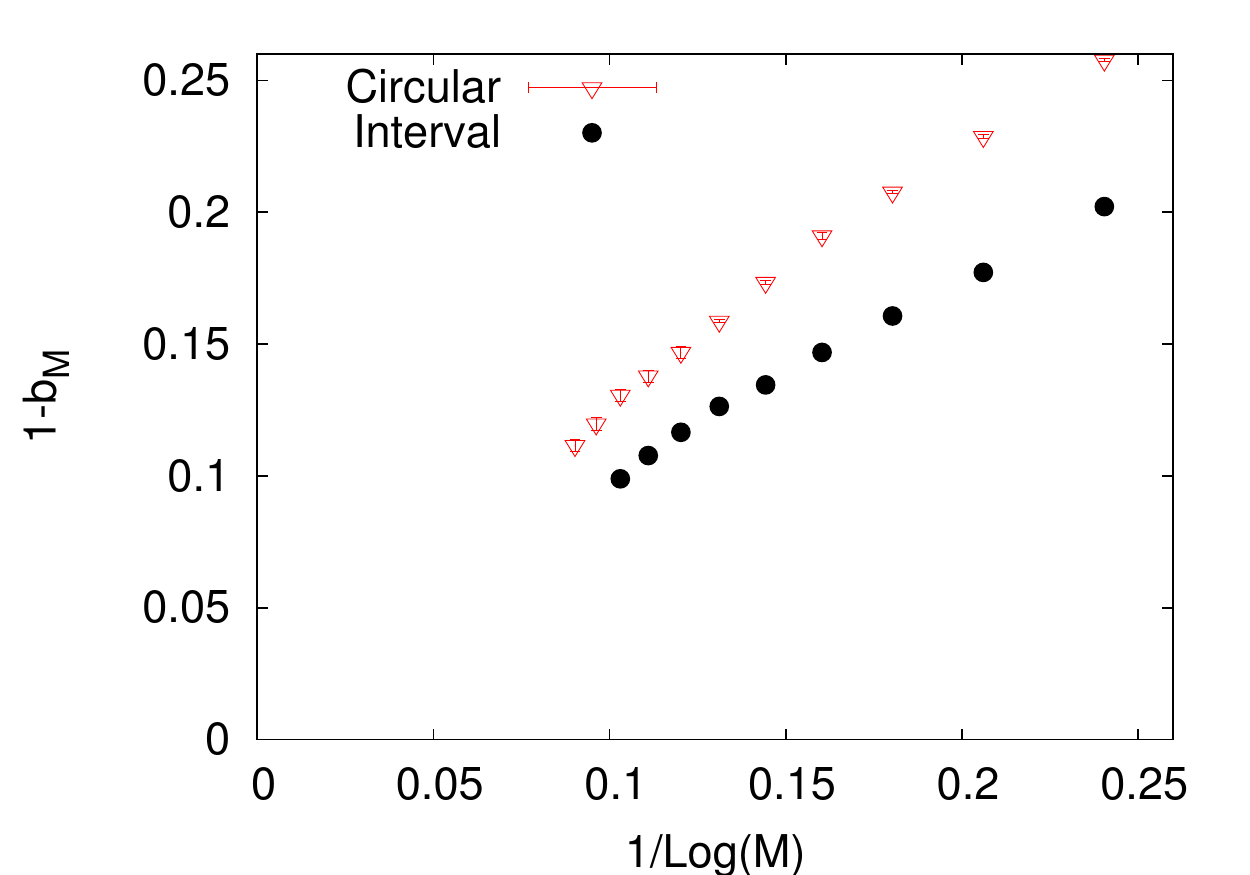}
  \caption{Color on line. Left: Finite size scaling of $a_M$ for variables with logarithmic correlations, circular ensemble Eq. (\ref{circmat}), and interval Eq.  (\ref{intervalM}), from $M=2^{8}$ to $M=2^{19}$. The predicted slope is $\tilde \gamma=3/2$, numerically we find $\tilde \gamma=1.4 \pm 0.1$. This is compared with independent random variables, the standard uncorrelated REM, where the prediction is $\tilde \gamma=1/2$ as observed. Right: Finite size effect for $b_M$ for variables with the same correlations.  The data are consistent with a convergence as $1/\log M$ and extrapolate to $b_M=1 \pm 0.02$, consistent with the predicted value $b_M=1$ in each case (which means an unrescaled variance $\overline{V_{min}^2}^c$ in agreement with the prediction given in the text, in each case).}
\label{fig:a_M}
\end{figure}

\begin{figure}[htpb]
  \centering
  \includegraphics[width=0.7\textwidth]{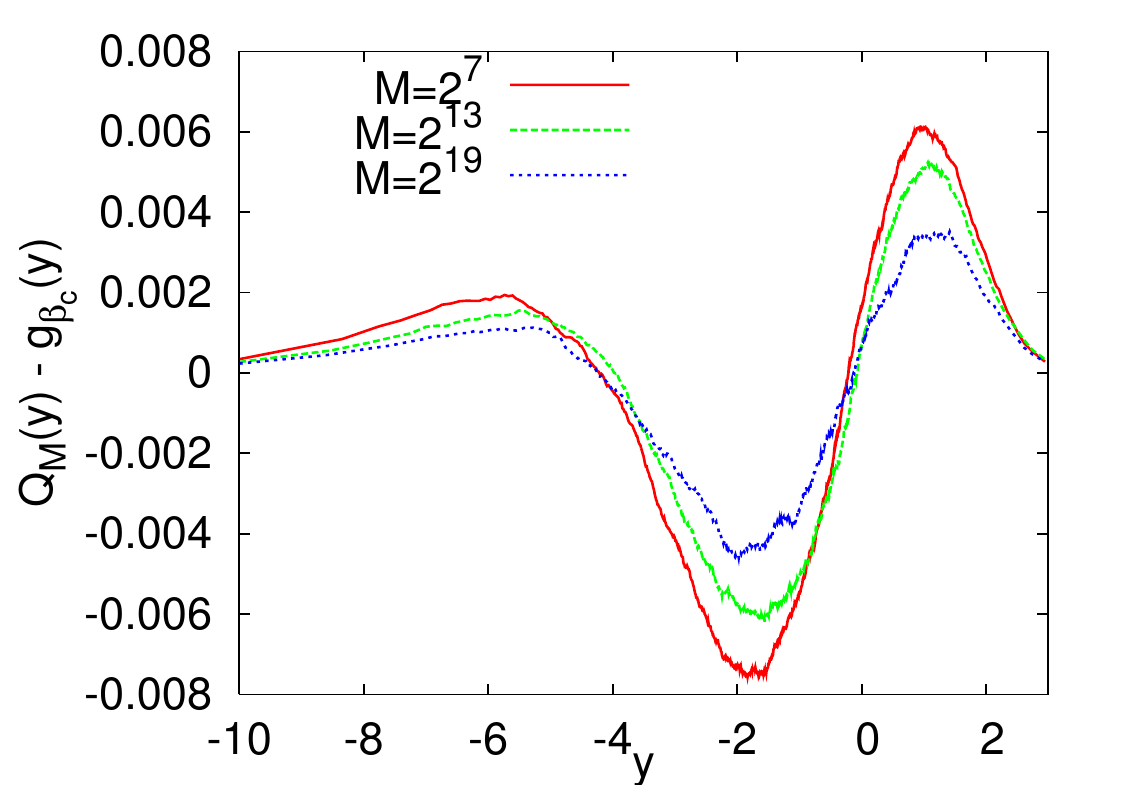}
  \caption{Color online. Circular case: cumulative distribution of the rescaled minimum $Q_M(y)$ minus the prediction (\ref{circ}) based on the freezing scenario $g_{\beta_c}(y)$. The number of samples is $10^7$.
The difference is small compared on the scale of unity. Although it is slow, the convergence is apparent.}
\label{fig:cumulative}
\end{figure}

\begin{figure}[htpb]
  \centering
  \includegraphics[width=0.7\textwidth]{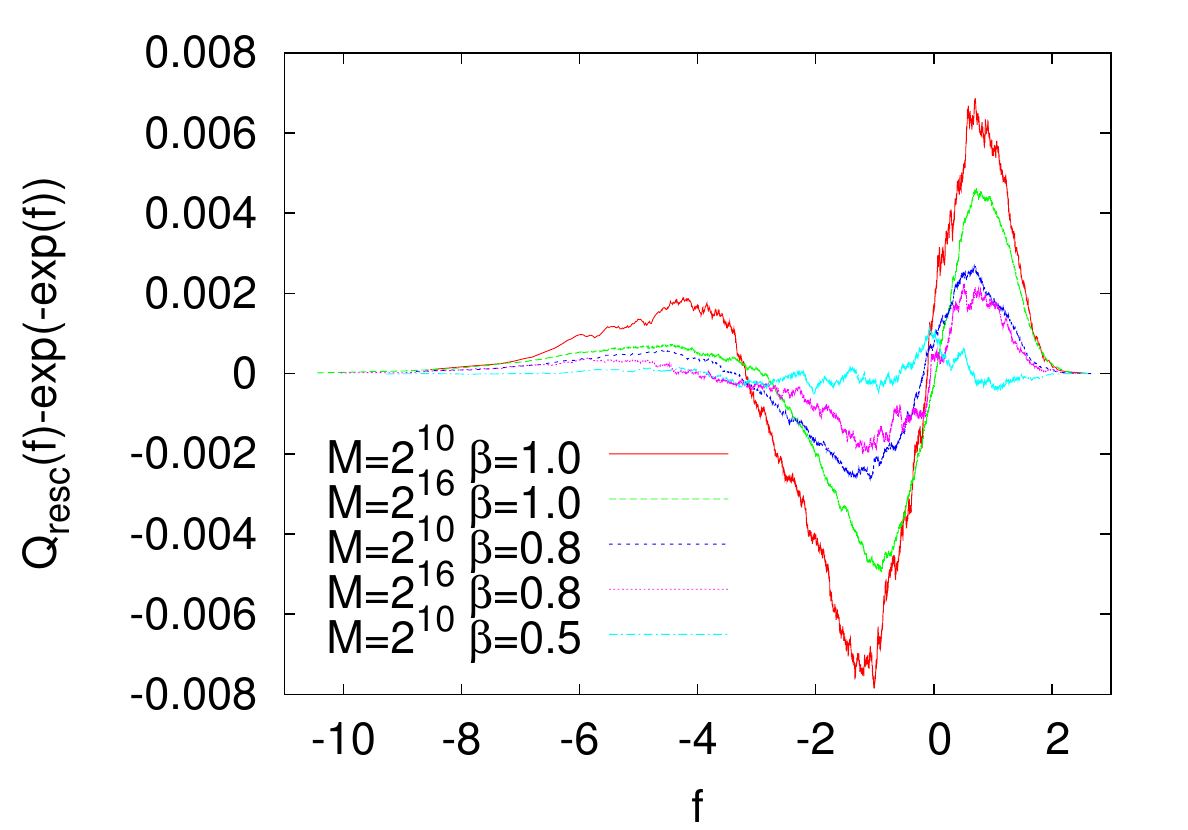}
  \caption{Color online. Circular case: distribution of the free energy in the high temperature phase, for various temperatures.}
\label{fig:freeenergy}
\end{figure}

From the distribution of the minimum $V_{min}$ in systems of up to $M=2^{19}$ we have computed the coefficients $a_M$, $b_M$ and the distribution of the variable $y$ in (\ref{rescaledmin}) by fixing $\overline{y}$ and the variance $\overline{y^2}^c$ to their value for the distribution (\ref{circ}). The asymptotics of the coefficients $a_M$ and $b_M$ in (\ref{rescaledmin}) are shown in Fig. \ref{fig:a_M}. They exhibit a reasonable agreement with the conjecture  (\ref{am&bm}) with $A=1$
but one clearly sees that convergence is slow. Convergence to $b_M=1$ would mean that the prediction $\overline{V_{min}^2}=\pi^2/3$ is correct. The cumulative distribution $Q_M(y)$ of the rescaled minimum, i.e. the variable $y$, is shown in Fig. \ref{fig:cumulative} where the cumulative distribution (\ref{circ}) has been substracted. One sees that although the difference is small its convergence, if any, to zero is extremely slow (empirically
a $\sim 1/\sqrt{\ln M}$ seems to roughly account for the data, but we do not wish to make any strong claim here).

Then we computed the distribution of the free energy at various temperatures. In Fig. \ref{fig:freeenergy} we have first normalized the free energy distribution to the same average and variance as the unit cumulative Gumbel distribution, i.e. $\exp(-e^x)$, then plotted the
difference between the resulting cumulative distribution $Q_{resc}(f)$ and the Gumbel expression. This shows that the convergence is very fast at $\beta=1/2$ but rather slow already at $\beta=1$, where we have little doubt for the result. This is consistent with the fact that the convergence for the minimum is so slow.

\begin{figure}[htpb]
  \centering
  \includegraphics[width=0.7\textwidth]{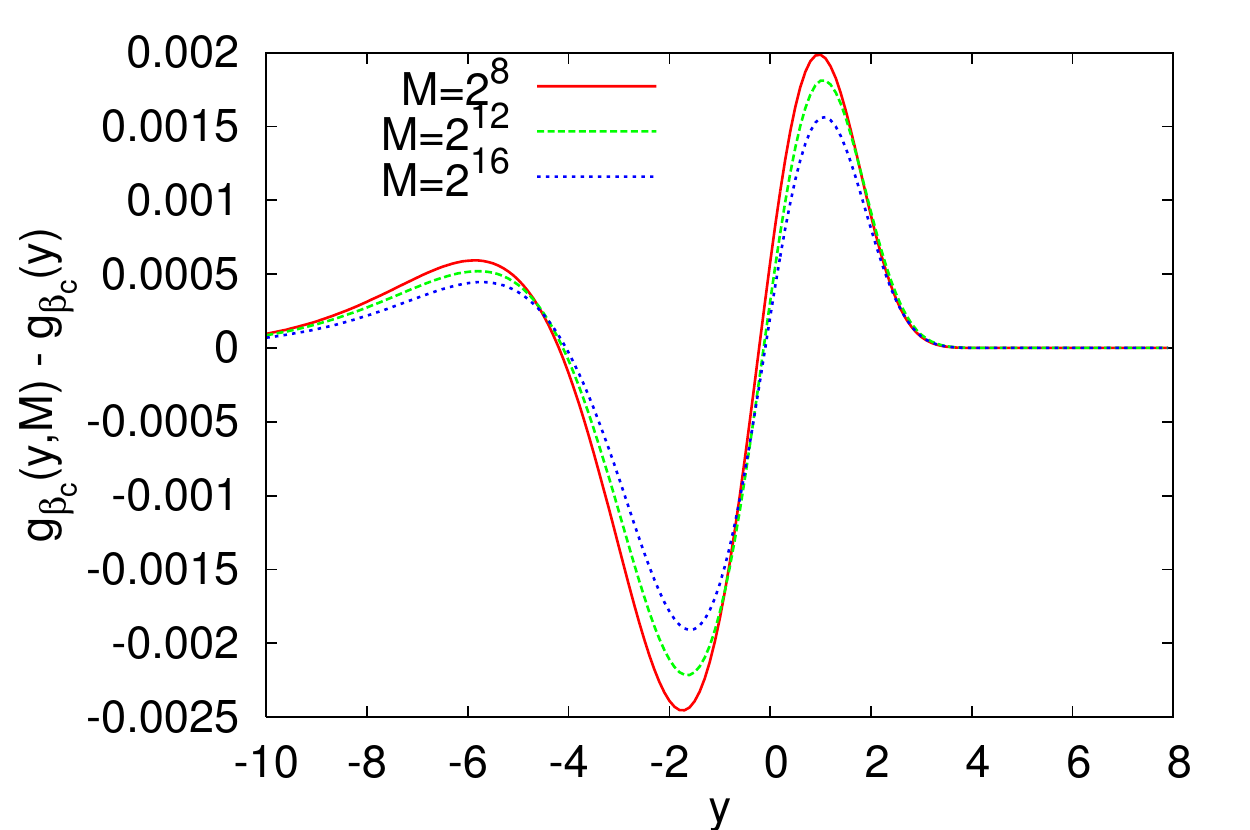}
  \caption{Color online. Circular case: convergence of $g_\beta(y,M)$ at $\beta=\beta_c$. We see that the scale is smaller than on Fig. 2 but that convergence is very slow.}
\label{fig:freeze1}
\end{figure}

\begin{figure}[htpb]
  \centering
  \includegraphics[width=0.7\textwidth]{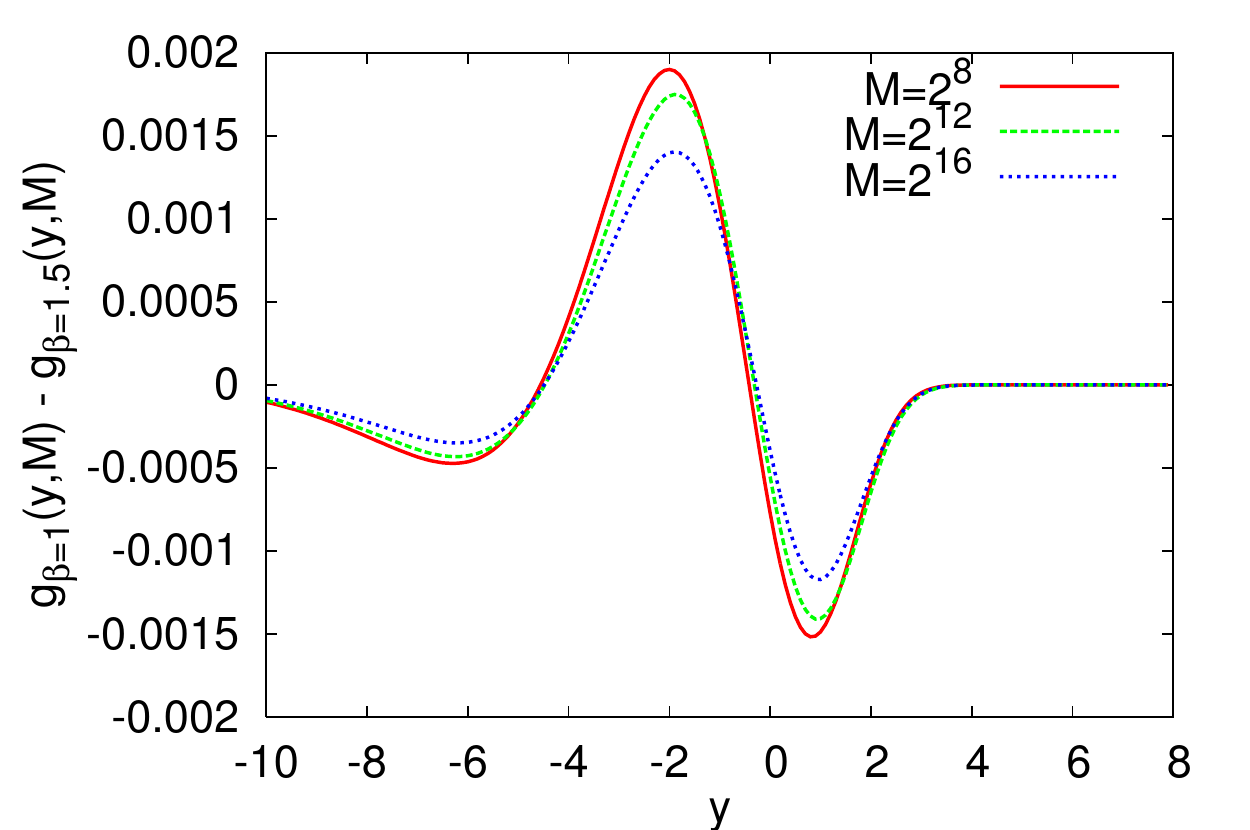}
  \caption{Color online. Circular case: direct test of the freezing scenario: convergence of $g_\beta(y,M)-g_{\beta_c}(y,M)$ (both are numerically measured and rescaled as explained in the text). We see that the scale is smaller by a factor around $2$ to $4$ than on Fig. \ref{fig:cumulative}, but that convergence is very slow.}
\label{fig:freeze2}
\end{figure}

To test the freezing scenario we also compute numerically $g_\beta(y)$ for various temperatures. First in Fig. \ref{fig:freeze1}
we test the convergence of the numerically determined $g_{\beta_c=1}(y,M)$ to the analytical prediction $g_{\beta_c}(y)$ in (\ref{circ})
as a function of $M$. Then in Fig. \ref{fig:freeze2} we
test whether $g_{\beta}(y,M)-g_{\beta_c}(y,M)$ at fixed $\beta>\beta_c$ decreases to zero as $M$ becomes large, which is the freezing conjecture. In practice we first compute the free energies $f_i$, compute their mean $\bar f$ and variance $\sigma$, define rescaled energies $f'_i=(f-\bar f +\gamma_E T - 2 \gamma_E)\sqrt{\frac{\pi^2}{3}(1-T^2/2)}/\sqrt{\sigma}$ and define $g_\beta(y,M)$ as the mean of $e^{-e^{\beta (y-f'_i)}}$ which, by construction and virtue of (\ref{cum}), has then the same average, $-2 \gamma_E$, and variance, $\pi^2/3$, as $g_{\beta_c}(y)$ in (\ref{circ}). Comparing Fig. \ref{fig:cumulative} and Fig. \ref{fig:freeze1} we see that a good fraction of the difference in Fig. \ref{fig:cumulative} is already due to finite size corrections at $\beta_c$ (which have nothing to do with the
testing the freezing scenario).

\subsection{universality of circular ensemble: cyclic matrices, GFF inside a disk
with Dirichlet boundary condition}

It is important to discuss now the universality of this result, as it is a rather subtle point. The general issue
of universality for logarithmic REM's can be formulated as follows. Consider sequences of $M$-dependent correlation matrices $C_{ij}^{(M)}$. What are the possible universality classes for the associated REM in the limit $M \to + \infty$, what are their basin of attraction and conditions for convergence? One may ask two questions: (i) extremal universality classes, i.e. correlation matrices which have asymptotically the same distribution of the minimum $V_{min}$ (up to a shift by a $M$-dependent constant) (ii) more restrictive universality classes valid for any $\beta$, i.e. correlation matrices which have asymptotically the same distribution of free energy, and generating function $g_\beta(y)$ (up to a shift by a $M$-dependent constant) for any $\beta$. It is reasonable to expect each latter universality class (ii) should correspond to a continuum model. Obviously, two sequences $C_{ij}^{(M)}$ which belong to the same class (ii) also have the same distribution of extrema. But there are counter examples to the reverse (see below). Classifying these classes being a formidable problem, here we only make a few remarks about the universality class of the circular ensemble. The class corresponding to the interval is discussed below.

Let us start from (\ref{circmat}) and discuss various generalizations in the subset of cyclic (called also periodic or circulant) matrices, i.e which can be written as:
 \begin{equation} \label{eigen}
 C_{ij}=\frac{1}{M} \sum_{k=0}^{M-1} \lambda_k e^{2 i \pi (i-j) \frac{k}{M}}.
 \end{equation}
with ($M$-dependent) real eigenvalues $\lambda_k$. The eigenvalue $\lambda_0$ corresponds to the uniform mode (often called zero mode in the GFF context). Logarithmic correlations mean that we assume that $\lambda_k \sim 1/k$ in some broad range of $k$ at large $M$, as specified below.

Starting from (\ref{circmat}) let us first make the observation that adding a fixed $W>0$ of $O(1)$ on the diagonal of $C_{ij}$ shifts all eigenvalues by a constant $O(1)$ and does not change the universality class, in both sense (i) and (ii), at large $M$. On the other hand, a shift:
 \begin{equation}
 C_{ij} \to C_{ij} + \sigma
 \end{equation}
for all $(i,j)$ shifts only the the uniform mode $\lambda_0 \to \lambda_0 + \sigma$. It is equivalent to add a global random gaussian shift $v$ to all $V_i$, i.e. $V_i \to V_i + v$ where $\sigma=\overline{v^2}$. It thus results in the
the convolution of the distribution of $V_{min}$ (and of the free energy) by a gaussian of variance $\sigma$. One such example, discussed again below, is to consider the distribution of the GFF (using the full plane Green function) on a circle of radius $R<1$ (and cutoff $R \epsilon$, i.e performing a global contraction): it shifts all $C_{ij} \to C_{ij} - 2 \ln R$ in (\ref{circmat}). Hence we keep in mind that there is really {\it a family of distributions differing by their second cumulant}, and will enforce in our numerics the condition $\lambda_0=0$ which we believe selects the distribution (\ref{circ}).

\subsubsection{GFF along an arbitrary circle}

\begin{figure}[htpb]
  \centering
  \includegraphics[width=0.5\textwidth]{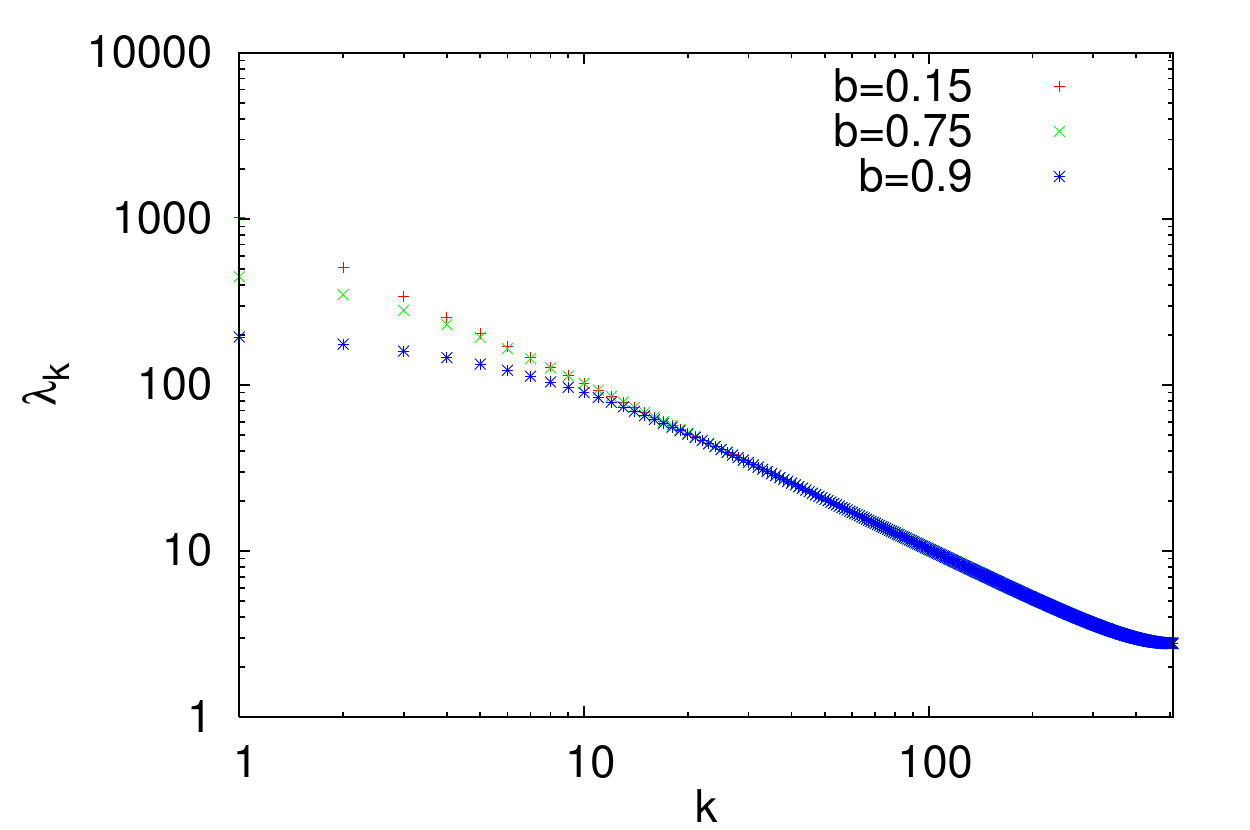}
  \caption{Color online. Eigenvalues of the correlation matrix corresponding to the
  GFF along a circle in a disk domain
  with zero boundary condition (Dirichlet). }
\label{fig:disk}
\end{figure}

One possible generalization of the circular model (\ref{circmat}) along these lines is the GFF inside a disk of radius $L$ with $V=0$ on the boundary as studied by e.g. Duplantier and Sheffield \cite{Qgrav}. Using the Dirichlet Green function $G_L(z,z') = - \ln \frac{ L |z-z'|}{|L^2 - z \bar z'|}$, the correlation matrix for the discrete model on a circle of radius $R$ inside the disk is then for $i \neq j$ and denoting $\rho=R/L$:
 \begin{eqnarray} \label{dirichlet}
&& C_{ij}  =  - 2 \ln \frac{ 2 \rho |\sin(\frac{\theta_i - \theta_j}{2})|}{\sqrt{1+\rho^4 - 2 \rho^2 \cos(\theta_i-\theta_j)}} \quad , \quad C_{ii} = 2 \ln L + 2 \ln (1- \rho^2) - 2 \ln \epsilon
 \end{eqnarray}
In the small $\rho=R/L$ limit, equivalently fixed $R$ and large $L$ one finds $C_{ij} \approx  -2 \ln \rho - 2 \ln ( 2 |\sin(\frac{\theta_i - \theta_j}{2}) |$ and $C_{ii} \approx 2 \ln L - 2 \ln \epsilon$. Choosing \cite{footnote} $\epsilon=R/M$ one sees that one recovers indeed the FB model (\ref{circmat}) (with $W=0$) {\it up to a shift $\sigma = 2 \ln (L/R)$ in the zero mode} $\lambda_0$ of the matrix, i.e all eigenvalues of the correlation matrix are the same as FB except the uniform mode. This gives us the precise meaning of the universality of the results of FB \cite{FB}: it holds for small $\rho=R/L$ for the Dirichlet GFF on the disk and up to a (trivial) convolution by a gaussian of width $2 \ln (L/R)$. The next question is whether the universality extends to other circular contours on the disk with $R/L$ not necessarily small. The answer is no, as can be argued from examination of the eigenvalues, diagonalizing (\ref{dirichlet}) for arbitrary $\rho$. As shown in Fig. \ref{fig:disk} at large $M$ the eigenvalues are essentially the same as the ones for FB, i.e. small $\rho$, apart from the few largest ones - whose number does not change and remains finite as $M$ becomes large. We expect that however that since these are the {\it largest} eigenvalues, despite being few they will change the distribution of the maximum which hence will depend continuously on the ratio $R/L$ (with a similar discussion as above concerning the zero mode and convolution by a gaussian).

\subsubsection{other periodic models}

\begin{figure}[htpb]
  \centering
  \includegraphics[width=0.5\textwidth]{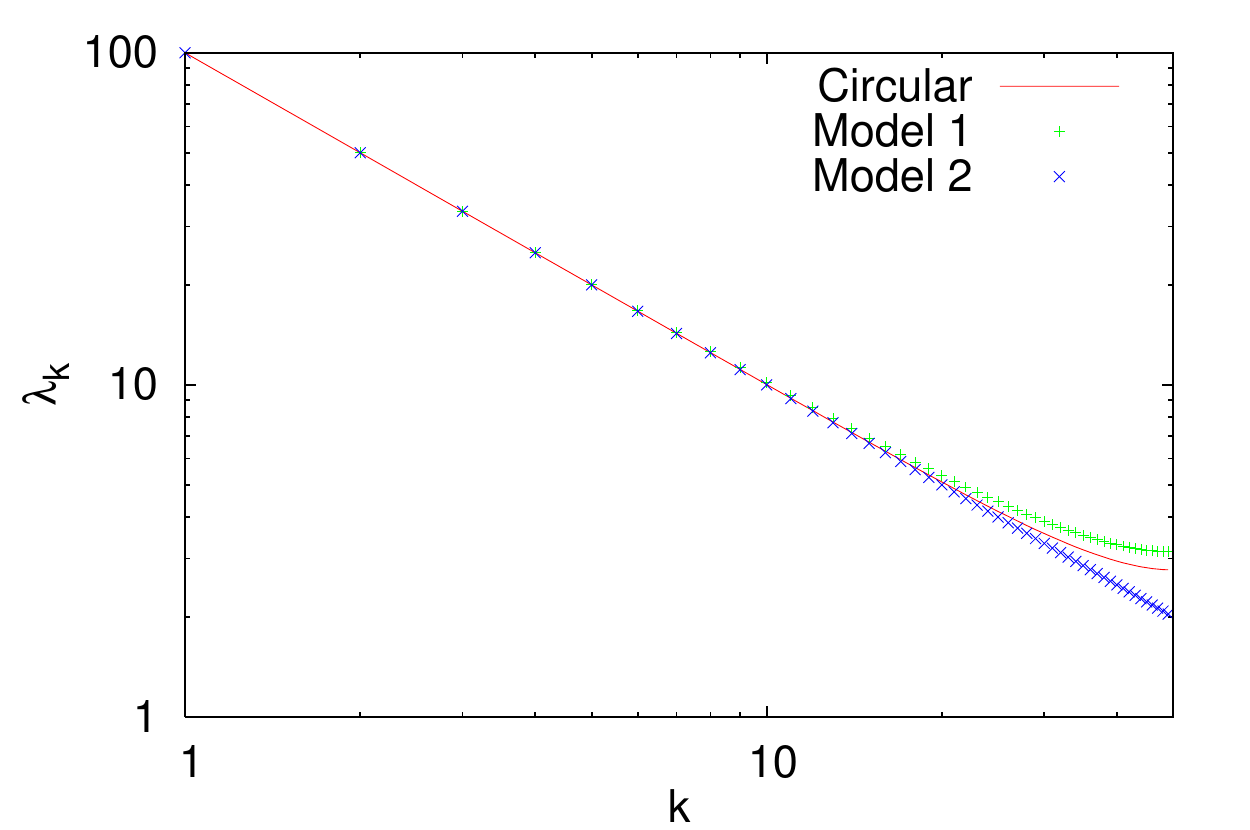}
  \caption{Color online. Eigenvalues of the correlation matrix corresponding to the periodic models defined in the text.}
\label{fig:longrangelambda}
\end{figure}

\begin{figure}[htpb]
  \centering
  \includegraphics[width=0.7\textwidth]{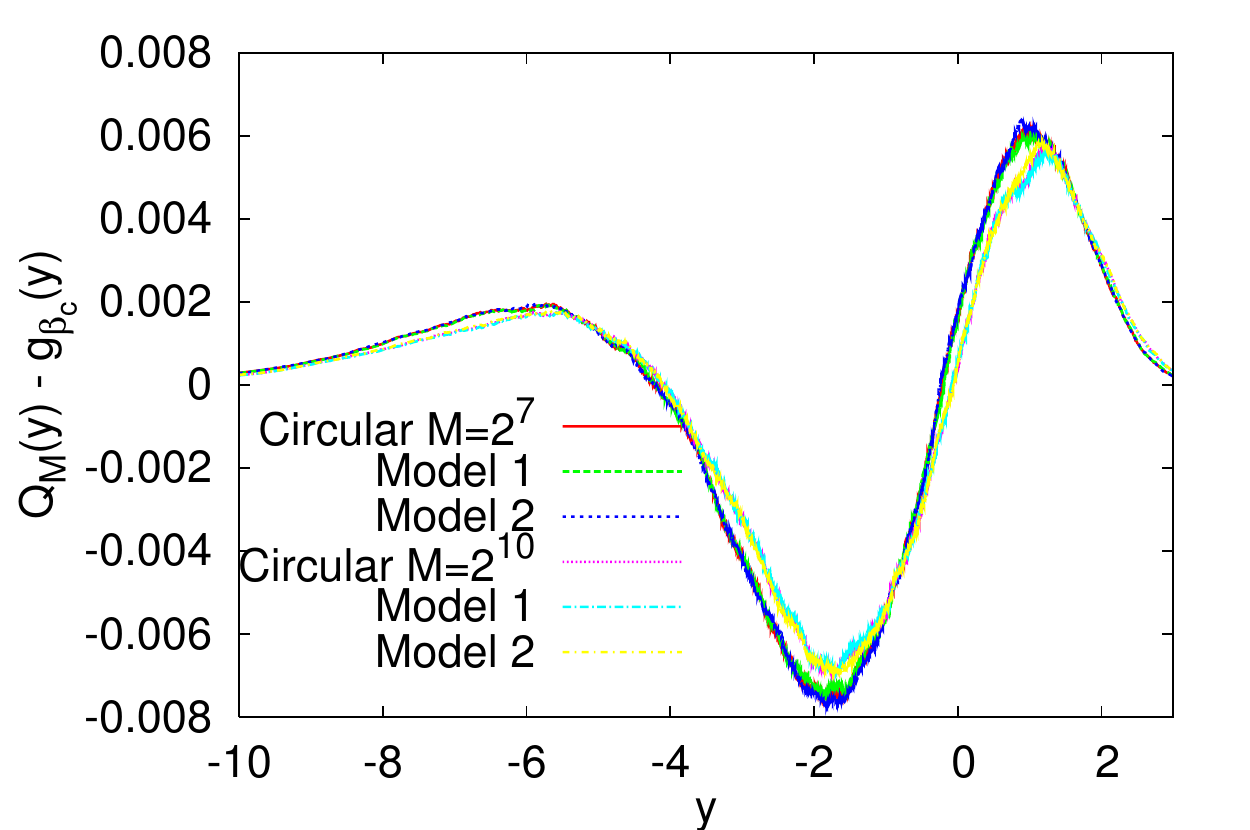}
  \caption{Color online. Universality in the case of periodic (circulant) correlation matrices: cumulative distribution of the minimum, with subtraction as in Fig. \ref{fig:cumulative}: convergence to a common curve is faster than to the global analytic prediction.}
\label{fig:longrange}
\end{figure}

On the other hand a much stronger universality property appears to hold when only the {\it smallest} eigenvalues are changed. Hence we now
test whether the results obtained for the circular case remain valid for all periodic cases (\ref{eigen}) with the same behaviour of $\lambda_k \sim 1/k$. Again it is important that $\lambda_0$ be fixed to zero. If $\lambda_0 >0$ this amounts to convoluting the distribution of the minimum by a Gaussian of variance $\lambda_0$. For the model to be logarithmic and strong universality to hold we require that $\lambda_k \to 1/k$ as $M \to \infty$ for $0<k \ll M$. We have tested this conjecture for two models.

Model 1: Sharp model (SM)  $\lambda_k=M/k$ for $k=1,\ldots, M/2$ and   $\lambda_k= M/(M-k)$ for $k=M/2,\ldots, M-1$

Model 2: The long range model (LRM) which is some discretization of the Joanny-deGennes elasticity of the contact line \cite{joanny}, $\lambda_k = 2 \pi/\sqrt{2(1-\cos(2 \pi k/M))}$.

The eigenvalues of these models are compared to the one of the circular case in Fig. \ref{fig:longrangelambda} and one can see that they differ only for $k$ near $M/2$. As can be seen in Fig. \ref{fig:longrange} the convergence of these models to the circular case at fixed $M$ is much faster than their (common) convergence to the analytical prediction. We take this as a signature of the strong universality with respect to variations of the correlation matrix which change only the smallest eigenvalues, within the cyclic class. We check in the
Appendix C that the first terms in the expansion of $\overline{f^2}^c$ are the same for all these models which
supports that the universality holds at any $\beta$, i.e. both in the sense (i) and (ii) defined above.

\subsection{interval}

We now discuss the $[0,1]$ ensemble. We take for the correlation matrix the Toeplitz form $C_{ij}=C(i-j)$, $i,j=1,..M$:
\begin{equation} \label{intervalM}
C_{ii}=  4\sum_1^{M-1}(-1)^k\log\frac{k}{M}  + W \;\;\; C_{i \neq j}= - 2 \log\frac{|i-j|}{M}
\label{choice}
\end{equation}
with $W=0$. This matrix is not diagonal in Fourier space and we cannot use the FFT method. In practice we find the eigenvalues $\lambda_k$ and the normalized eigenvectors $\psi_k(i)$ by  a direct diagonalization of the matrix $C_{ij}$. We then generate the correlated random potential as $V_i=\sum_{k=0}^{M-1}\sqrt{\lambda _k} x_k  \psi_k(i)$, where the $x_k$  are  i.i.d. real unit centered Gaussian variables. Performing this sum together with the direct diagonalization is numerically expensive and limits the size of the number $M$ of correlated numbers.  In order to achieve a good statistics ($\sim 10^7$ samples) we analyze here data only up to $M=2^{12}$.

To justify our choice for the diagonal element in (\ref{intervalM}) let us recall a useful property of any Toeplitz matrix: if the function $f(\theta)=C_{11}+2\sum_{k=1}^{M-1}C_{1,k+1} \cos{(k\theta)}$
is positive $\forall \, \theta\in[0,2\pi)$, then $C$ is positive definite (for any $M$). This is seen by noting that for any vector $v_k$, $k=0,..M-1$, one has $\sum_{k,\ell=0}^M v_k v_{\ell} C(k-l) = \int_0^{2 \pi} \frac{d\theta}{2 \pi} f(\theta) |v(\theta)|^2$ where $v(\theta)=\sum_{k=0}^{M-1} v_k e^{i \theta k}$ and
$C(k)=\int_0^{2 \pi} \frac{d\theta}{2 \pi} f(\theta) e^{i k \theta}$. More importantly it can be shown
that the reverse is true for large $M$ \cite{toeplitz}. For the choice of Eq(\ref{choice}) this function has a global minimum at $\theta =\pi$, for which $f(\theta=\pi)=0$. As a result the matrix $C_{ij}$ is positive definite and in the large $M$ limit, one finds that the smallest eigenvalue goes rapidly to zero and the eigenvector components alternate as $(-1)^k$. Note that the diagonal element in (\ref{intervalM}) behaves as $C_{ii} \sim 2 \ln M + O(1)$ at large $M$, hence as expected, and similarly to the circular case. Though a convenient choice to prove positivity, there are other choices with similar behaviours at large $M$ which would do as well.

\begin{figure}[htpb]
  \centering
  \includegraphics[width=0.7\textwidth]{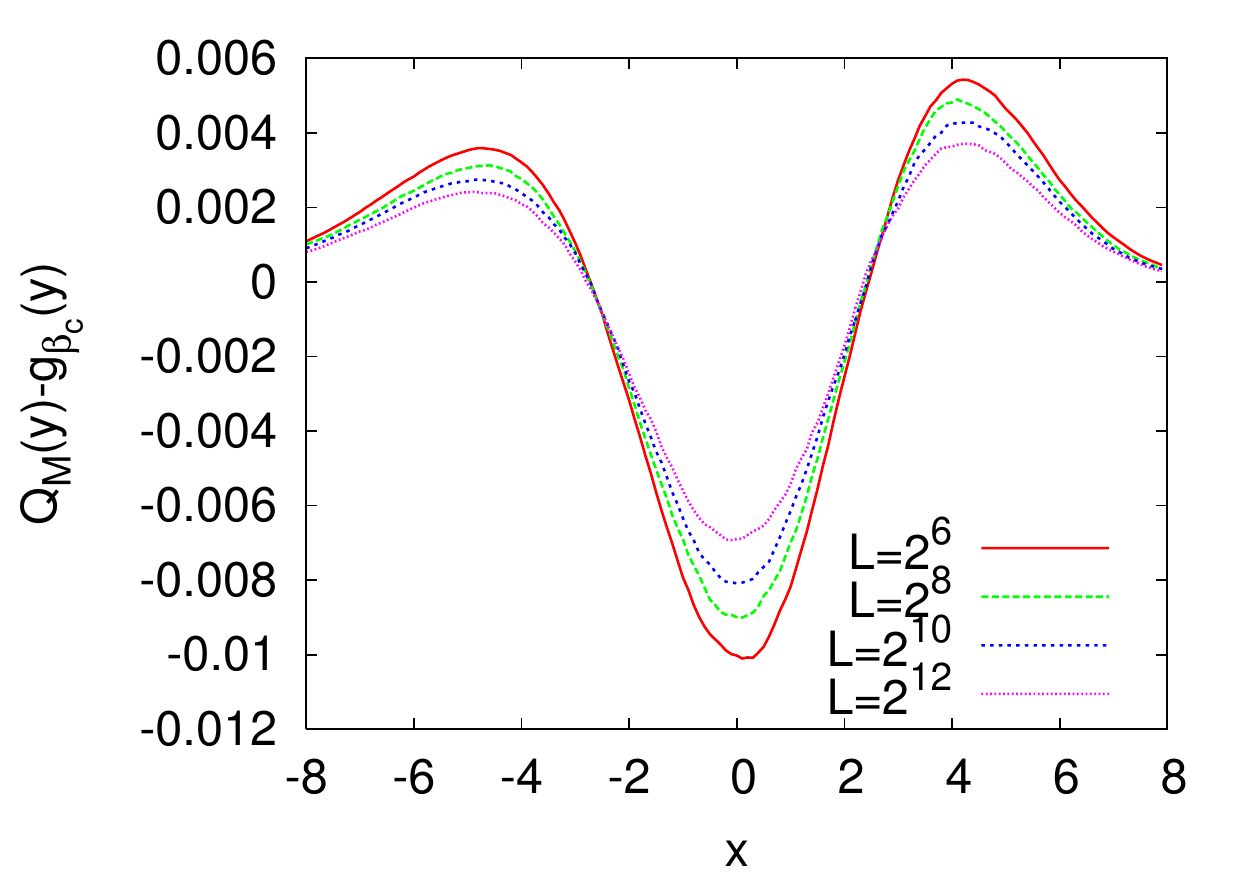}
  \caption{Color online. Interval case: cumulative distribution of the rescaled minimum $Q_M(y)$ minus our analytical prediction, $g_{\beta_c}(y)$,  shown in Fig. \ref{fig:gfree} and based on the freezing scenario. The number of samples is $10^7$.
The difference is small compared on the scale of unity. Although it is slow, the convergence is apparent.}
\label{fig:cumulinterval}
\end{figure}

We have analyzed the distribution of the minimum $V_{min}$ and computed the
 coefficients $a_M$, $b_M$ and the distribution of the variable $y$ in
 (\ref{rescaledmin}) by fixing $\overline{y}=7/2 -2 \gamma_E - \ln(2 \pi)$ and the variance $\overline{y^2}^c=\frac{4}{3} \pi^2 - \frac{27}{4}$ to their value given by the analytically prediction.
  The convergence to $b_M=1$, shown in Fig. \ref{fig:a_M} (right), is thus a test of our prediction $\overline{V_{min}^2}^c=\frac{4}{3} \pi^2 -
 \frac{27}{4}$. The convergence of the coefficients $a_M$ and
 $b_M$ is quite similar to the circular case. The cumulative distribution
 $Q_M(y)$ of the rescaled minimum, i.e. the variable $y$, is shown in Fig.\ref{fig:cumulinterval} where the cumulative distribution of Fig.\ref{fig:gfree} (our analytical prediction) has been substracted. Again, the behaviour resembles the one for the circular case.

The discussion of the universality for the interval class is more delicate since now the lowest eigenvector is no more generically the uniform mode. However a way to realize it from the GFF can be suggested similarly to the above discussion. One can consider the interval embedded near the center in a large disk with Dirichlet b.c. In the limit of small ratio $\rho$ of interval size to disk radius the above interval model applies, again up to a convolution by a gaussian of variance $2 \ln(1/\rho)$.


\bigskip

\subsection{temperature dependence of the second cumulant of the free energy}

\begin{figure}[h]
  \centering
  \includegraphics[width=0.45\textwidth]{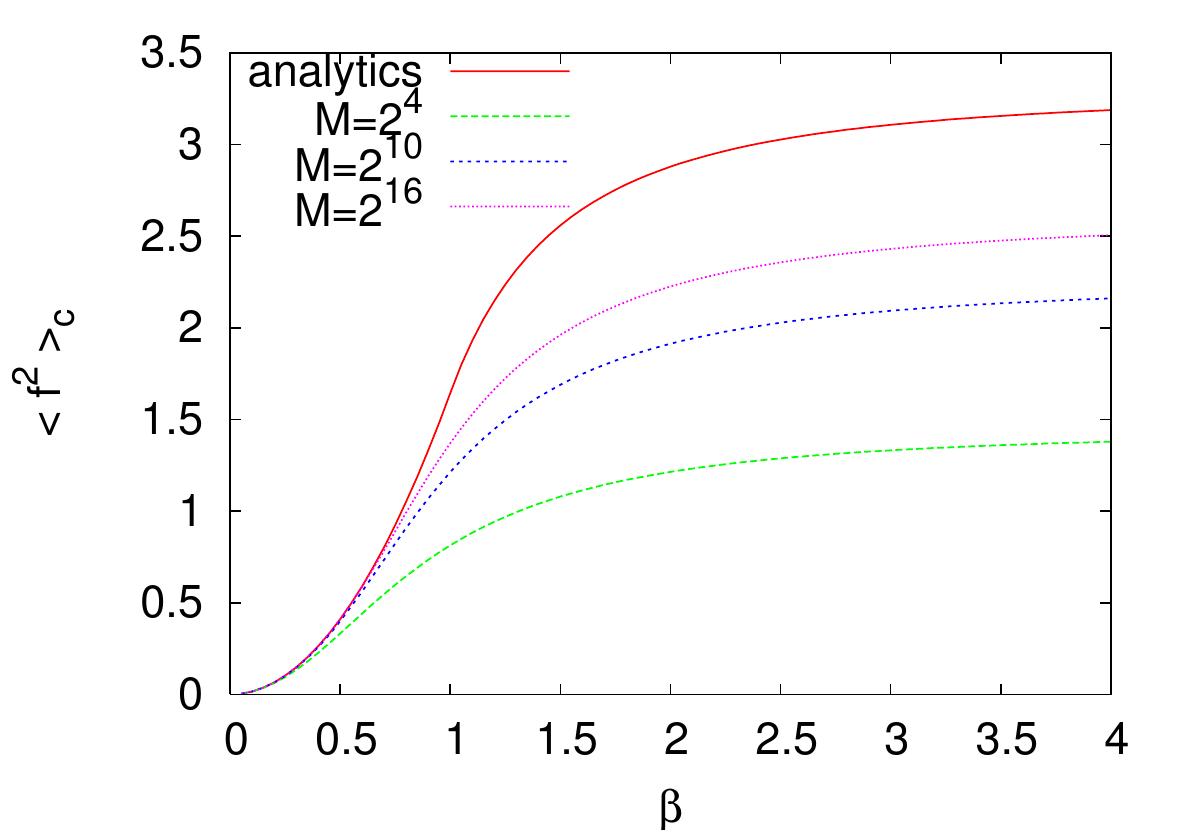} \includegraphics[width=0.45\textwidth]{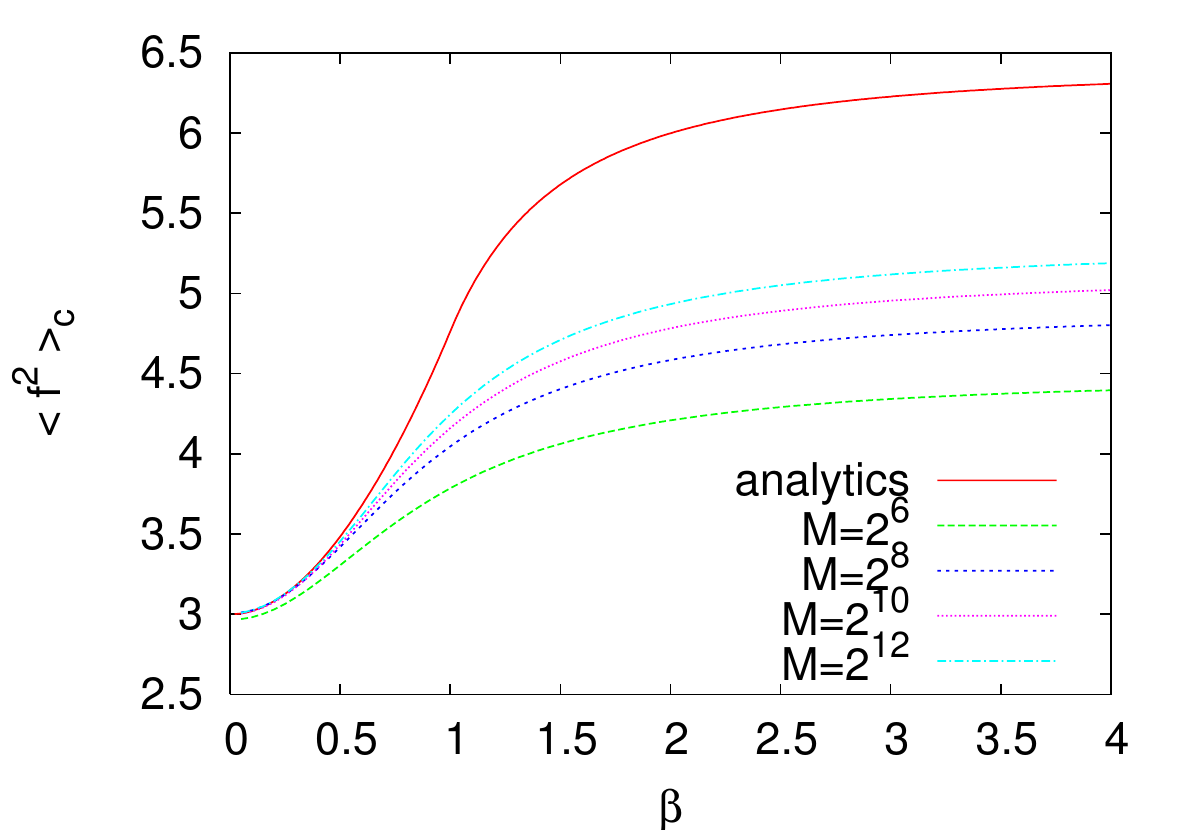}
  \caption{Color online. Second cumulant $\overline{f^2}^c$ of the free enenergy as a function of inverse temperature $\beta$ for various sizes, as compared to the analytical prediction given in the text.
  Left: circular ensemble. Right: interval case. }
\label{fig:f2}
\end{figure}

\begin{figure}[h]
  \centering
  \includegraphics[width=0.45\textwidth]{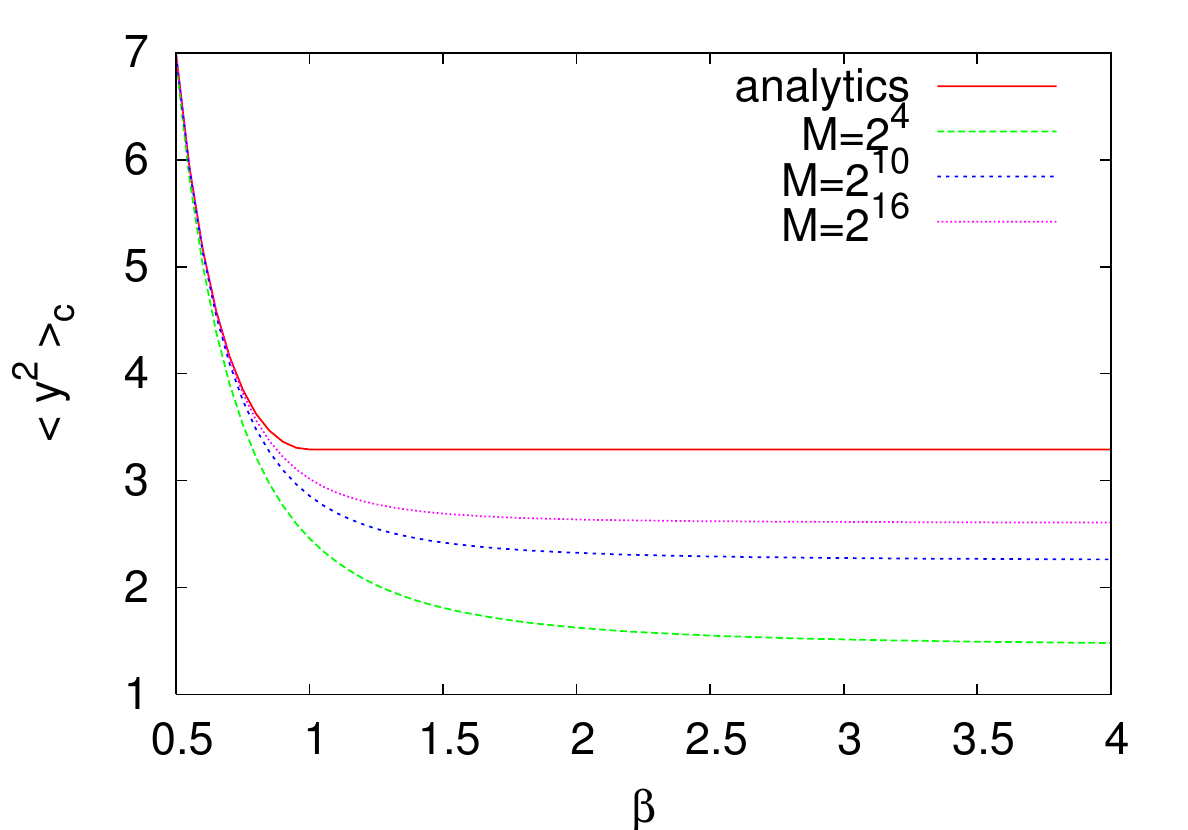}
  \caption{Color online. Circular case: second cumulant $\overline{y^2}^c$ as a function of inverse temperature $\beta$ for various sizes $M$ as compared to the analytic prediction given in the text.}
\label{fig:y2}
\end{figure}

Finally we have also performed some numerical tests of the temperature dependence of our analytical results in the high temperature phase. We have computed numerically, and plotted in Fig. \ref{fig:f2} and \ref{fig:y2}  as functions of $\beta$, the variance of the free energy distribution $\overline{f^2}^c$ as well as $\overline{y^2}^c$ for the circular case (\ref{circmat}) and $\overline{f^2}^c$ for the interval case (\ref{intervalM}). They are compared to the analytical predictions, i.e. (i) for the circular case:
\begin{eqnarray} \label{f2circ2}
\overline{f^2}^c  &=&  (\pi^2/6) \beta^2    \quad (\beta<1)  \quad {\rm and} \quad \overline{f^2}^c =  (\pi^2/6) (2 - T^2)  \quad (\beta>1)
\end{eqnarray}
which via Eq. (\ref{cum}) corresponds to $\overline{y^2}^c =  \frac{\pi^2}{6} ( \beta^2 + \frac{1}{\beta^2})$ for $\beta<1$ which freezes into $\overline{y^2}^c =  \pi^2/3$ for $\beta>1$, and, (ii) for the interval case
formula (\ref{f2int}) for $\beta<1$ and $\overline{f^2}^c(\beta) = \overline{f^2}^c(\beta_c=1) + \frac{\pi^2}{6}(1-T^2)$ for $\beta>1$. One can verify the good convergence in the high temperature phase. Questions related to the behavior for small $\beta$, and how the numerical convergence could be further improved, is discussed in the Appendix C.

\subsection{more open questions on universality}

Let us now indicating a simple example where universality (i) of distribution of minimum and
of (ii) the free energy at any temperature, discussed above, may differ from each other. Consider the continuum problem on the circle but
with an arbitrary smooth and a non singular weight $0<\rho_1 < \rho(\theta) < \rho_2 <1$:
\begin{eqnarray} \label{1prim}
&& Z= \epsilon^{\beta^2} \int_{0}^{2 \pi} d \tilde \theta \rho(\tilde \theta) e^{- \beta V(e^{i \tilde \theta})}
= \epsilon^{\beta^2} \int_{0}^{2 \pi} d \theta e^{- \beta V(e^{i f(\theta)})}
\end{eqnarray}
and we consider for instance $\tilde \theta=f(\theta)=\theta+ a \sin(\theta)$ with $a<1$ and
$\rho(\tilde \theta)=1/f'(\theta)$. From the second form in (\ref{1}) one sees that the associated REM
can be chosen as $Z_M=\sum_i e^{- \beta V_i}$ with correlation matrix $C_{ij} = - 2 \ln|2 \sin(\frac{1}{2} f(\theta_i)-\frac{1}{2} f(\theta_j))|$ for $i \neq j$ and $\theta_i=2 \pi i/M$, neither a circulant nor Toeplitz matrix.
As shown in the Appendix C, at small $a$, $\overline{f^2}^c=\frac{1}{2} a^2 + O(a^4) + O(\beta^2)$, hence the free energy distribution clearly depends on $a$. On the other hand, the first form in (\ref{1}) suggests that it should have the same distribution of the minimum $V_{min}$ as $\rho(\theta)=1$.  Indeed,
$V(\theta)-T\ln \rho(\theta)$ as $T \to 0$ should have same extremal statistics for a given $\epsilon$ regularization (with $T \to 0$ before $\epsilon \to 0$) as $V(\theta)$ provided $\rho(\theta)$ is non singular. The question of how the freezing scenario works under such a circumstances and what are the universality classes is left for future studies.

Finally, another challenging question about universality, related to the GFF, is about REM's constructed along different curves in the plane than the circle, or the interval, i.e. $V_i=V(z_i)$ where the $z_i$ lie along a curve and sample it at large $M$ with a density described by some given arc length $\sqrt{d z d \bar z} \rho(z,\bar z)$. One can use conformal maps to relate various curves to each others, e.g. a circle to a slightly deformed circle, with different weight functions $\rho(z,\bar z)$. Hence we are back to understanding the type of problem described in the preceding paragraph, and one should expect some universality in the distribution of the minimum.

\section{Conclusion}

To summarize, we have studied analytically and numerically random energy models based on
Gaussian random potentials with logarithmic correlations. We have extended the Fyodorov-Bouchaud (FB) results from the circular ensemble to the interval. We have found the proper analytic continuation from the positive integer moments of the partition function, expressed as Selberg integrals, to arbitrary moments.
This analytic continuation of the Selberg integrals, previously an outstanding open problem, is solved here.
The solution involves Barnes functions and their generalizations which appear in studies of the Liouville field theory,
hence strengthening the already noted link between the two problems. This solution, valid in the high temperature phase,
allowed us to obtain the full distribution of the free energy $f$ for $\beta \leq  \beta_c$ and up to the critical point.
It was generalized, at $\beta=\beta_c$ to the case where additional charges exist at the end of the interval.

The knowledge of the generating function $g_\beta(y)=\overline{\exp(- e^{\beta (y-f)}) }$ at $\beta=\beta_c$ allowed us,
via the same {\it freezing scenario} hypothesis as put forward in FB for the circular ensemble,
to obtain the distribution of the minimum of the gaussian free field (GFF) on an interval,
expressed as an integral transform of a Barnes function. The freezing scenario,
which asserts that precisely this generating function $g_\beta(y)$ becomes temperature independent in the glass phase for $\beta \geq \beta_c$, was until now based on a traveling wave analysis. While rigorous for the Cayley tree based REM, for which it was introduced, it was only based on a one-loop RG analysis for the type of models at hand \cite{carpentier}. Here we made what we believe should be considered as a step towards better understanding of this freezing scenario: we discovered that, both for the circular ensemble and its interval counterpart, the analytic expression of $g_\beta(y)$ obeys in the high temperature phase the duality with respect to the transformation $\beta \to 1/\beta$. It implies in particular $\partial_\beta g_\beta(y)=0$, for all $y$, at $\beta=\beta_c$ in perfect agreement with a continuous freezing scenario. While one may notice that the generating function $g_\beta(y)$ is special as being the partition function of the Liouville {\it model} (see e.g. the discussion in \cite{carpentier}) further connections to duality in Liouville {\it field theory} remains to be understood (the high temperature phase being the analogous of the weak coupling phase in Liouville).

Detailed numerical calculations of the free energy distribution and of the function $g_\beta(y)$ associated to discrete REM versions of the circular and interval models were performed. The freezing scenario is consistent with our results, in both cases, though convergence is found to be very slow. The numerically obtained distribution of the minimum $V_{min}$ of $M$ random Gaussian variables $V_i$ with logarithmic correlation matrices $C_{ij}$ is found to lie close to the predictions, but with only very slow convergence as a function of $M$. In the high temperature phase the convergence to the FB result for the circular case and to the present one for the interval is found to be very convincing, and in full agreement with various high temperature expansions also performed here.

The important question of the universality classes for discrete REM based on logarithmic matrices $C_{ij}$ and for their continuum analogs, is discussed. The continuum circular ensemble of FB is found to provide a single universality class for all circulant matrices with appropriate behavior of their spectrum at large $M$, for which we provide several examples. This strong version of universality holds for any temperature, i.e. identical distribution of free energy for all $\beta$, up to a shift. A weaker version of universality, holding only for the distribution of the minimum $V_{min}$ is discussed through an example. As far as the connection to the GFF is concerned, we discuss the case where the field is sampled along a circle of radius $R$ inside a disk of radius $L$ with Dirichlet boundary condition. We demonstrate universality, {\it up to the convolution by a gaussian}, in the limit of a small ratio $R/L$, while in general the distributions depend on the aspect ratio $R/L$.

The present progress opens many more fascinating questions. First one would want to extend these results to other curves in the plane, and even to two dimensional regions. The simplest extension, i.e. the case of the real axis with gaussian weight, also studied here, and for which the present methods are found to fail, shows that more remains to be understood before this can be achieved. Unbounded regions seem to pose a problem, and so does the control of the zero mode. The question of classifying the universality classes remains as a tantalizing open question. One can expect that the conformal invariance of the 2d GFF will play a crucial role in that classification, as it allows to map one curve into another one, with a change in the local length element. The question of which models obey duality and what is the precise connection to the freezing scenario is also outstanding. Further exploration of the connection to the Liouville model, to the Liouville field theory and to Liouville quantum gravity measures, is an important direction for further research. In particular, the distribution of the length of a segment in Liouville quantum gravity seems to directly connect to our results.

{\it Acknowledgments}: We are grateful to I. Gruzberg and P. Wiegmann for useful
discussions at various stages of the project. YF acknowledges
support by the Leverhulme Research Fellowship project "A single
particle in random energy landscapes" and PLD from ANR program
05-BLAN-0099-01.

\medskip

{\it Note added:}

After submission, we learned of a recent independent study by D.
Ostrovsky \cite{Ostrovsky2009}  who obtained a high temperature
expansion of arbitrary moments for the $[0,1]$ problem with no edge
charges, and conjectured a formula for these moments. Exploiting the
integral representation (\ref{ZamoBarn}) for the generalized
Barnes function $G_\beta(z)$ together with the following doubling
formula:
\begin{equation}\label{doubling}
G_\beta(2z) = C_\beta  2^{2 z^2 - (1+ \beta + \frac{1}{\beta}) z} \pi^{-z}  G_\beta(z) G_\beta(z+\frac{1}{2 \beta}) G_\beta(z+\frac{\beta}{2}) G_\beta(z+ \frac{1}{2 \beta} + \frac{\beta}{2})
\end{equation}
where $C_\beta$ is determined from e.g. $z=1$, we were able to show
that his conjecture is equivalent to our formula (\ref{MT}).
 Note however that no discussion of the critical case, duality and freezing is given in \cite{Ostrovsky2009}.

We have also shown that using Dirichlet boundary conditions at large distance $|x|=L$ for the 2D GFF gives a proper meaning to the problematic Gaussian weight case (it yields a shift $2 ln L$ in the second cumulant $\overline{y^2}$, while maintaining all higher cumulants as given in the text).

\appendix

\section{the special case of $[0,1]_{-1/2.-1/2}$}

Here we study the model defined by the partition sum:
\begin{eqnarray}
&& Z = \int_{-1}^{+1} \frac{dx}{\sqrt{1-x^2}} e^{\beta V(x)} \\
&& \overline{V(x) V(x')} = - 2 \ln |x-x'|
\end{eqnarray}
which is a special case of the interval $[0,1]_{ab}$ defined in the text for
$a=b=-1/2$. We show that it corresponds to a REM with a correlation matrix
which can be diagonalized in the Fourier basis.

Using the change of variable $x=\cos \theta $, hence $dx=\sin \theta d \theta= \sqrt{1-x^2} d\theta$, we see
that $Z$ can as well be written as an integral over a half-circle, involving a new gaussian random potential
with a modified correlator:
\begin{eqnarray}
&& Z =  \int_{0}^{ \pi} d \theta e^{\beta \tilde V(\theta)} \\
&& \overline{\tilde V(\theta) \tilde V(\theta')} = - 2 \ln |\cos(\theta) - \cos(\theta')|
= \sum_{n=1}^\infty \frac{4}{n} \cos( n \theta) \cos(n \theta') + 2 \ln 2
\end{eqnarray}
where we have used the formula:
\begin{eqnarray}
&& \sum_{n=1}^\infty \frac{2}{n} \cos( n A) \cos(n B) = - \ln(2 |\cos A - \cos B|)
\end{eqnarray}

To define the corresponding REM we now take a grid $\theta=2 \pi i/M$ for {\it the full circle} and take for correlation matrix:
\begin{eqnarray}
&& C_{ij} = \sum_{n=1}^\infty \frac{4}{n} \cos( 2 n \pi i/M) \cos(2 n \pi j/M) + 2 \ln 2
\end{eqnarray}
for $i \neq j$. The sum is still infinite, but it has the nice property that it can be made finite.
Indeed using that:
\begin{eqnarray}
&& \cos( 2 (k + m M) \pi i/M) \cos(2 (k+ m M) \pi j/M) =  \cos( 2 k \pi i/M) \cos(2 k \pi j/M)
\end{eqnarray}
we can now rewrite:
\begin{eqnarray}
&& C_{ij} = \sum_{n=1}^{M} \lambda_n \cos( 2 n \pi i/M) \cos(2 n \pi j/M) + 2 \ln 2 \\
&& \lambda_n = \frac{1}{n} + \sum_{m=1}^\infty ( \frac{1}{n+m M} - \frac{1}{m M})
=  \frac{1}{n}  - \frac{\gamma + \psi(1+\frac{n}{M})}{M}
\end{eqnarray}
One can check numerically that all the eigenvalues are positive.
Note that we have subtracted an infinite part on the diagonal so now the diagonal
element is also well defined:
\begin{eqnarray}
&& C_{ii} = \sum_{n=1}^{M} \lambda_n \cos( 2 n \pi i/M)^2  + 2 \ln 2
\end{eqnarray}
Hence for this particular interval model $a=b=-1/2$ we can use the Fourier basis to
generate the variables on the full circle and take the minimum only for
the half circle (i.e. $M/2 \times M/2$ submatrix). It remains to be understood
how this links to the simplification observed in formula (\ref{simpler}) in the text, and whether there
are other examples of such cases where a Fourier basis can be used.

\section{Some properties of the generalized Barnes function}

Let us first check that the function $G_\beta(x)$ defined by (\ref{ZamoBarn}) does indeed satisfy the property  (\ref{Gt1}).
We start from the formula (see \cite{GR} 8.341.3, p.889):

\begin{equation}\label{loggamma}
 \ln  \Gamma(x \beta) = \int_0^\infty \frac{dt}{t} \big[ \frac{e^{- \beta x t} - e^{- t}}{1-e^{-t}}
+e^{-t}( \beta x-1) \big]
\end{equation}

Now, by straightforward algebra (\ref{ZamoBarn}) implies:
\begin{eqnarray}\label{ZamoBarn1}
\! \! \! \! \! \! \! \! \! \! \! \! \! \! \! \!  \ln G_\beta(x+\beta)-\ln G_\beta(x) = \frac{\beta}{2} \ln(2 \pi) +  \int_0^\infty \frac{dt}{t} \big[ \frac{e^{- x t}}{(1-e^{-t/\beta})}
+\frac{e^{-t}}{2}(2\beta x-1) -\frac{\beta}{t} \big]
\end{eqnarray}
Changing now $t\to  \beta t$, and subtracting (\ref{loggamma}) gives:
\begin{eqnarray}\label{ZamoBarn2}\fl
&&  \phi_\beta(x)\equiv \ln G_\beta(x+\beta)-\ln G_\beta(x)-\ln \Gamma(x \beta) - \frac{\beta}{2} \ln(2 \pi)  \\
 && = \int_0^\infty \frac{dt}{t} \big[ \frac{e^{- t}}{1-e^{-t}}
-\frac{e^{-t \beta}}{2} + \beta x\left(e^{-t \beta}-e^{-t}\right)+ e^{-t}-\frac{1}{t} \big]
\end{eqnarray}
Now using the identity
\begin{equation}\label{log}
\int_0^\infty \frac{dt}{t}\left(e^{-t}-e^{-t \beta}\right)=\ln{\beta}
\end{equation}
we see that $\frac{\partial}{\partial x}\phi_\beta(x) =-\beta\ln{\beta }$
which implies $$\phi_\beta(x)=-\beta x \ln{\beta}+\phi_\beta(0),$$ with
\begin{eqnarray}\label{ZamoBarn2}
&& \phi_\beta(0) = \int_0^\infty \frac{dt}{t} \big( \frac{ e^{-t}}{1-e^{-t}}
-\frac{e^{-t \beta }}{2} + e^{-t}-\frac{1}{t} \big)
\end{eqnarray}
In turn, it is easy to see this integral converges and $\frac{d}{d\beta}\phi_\beta(0)=\frac{1}{2\beta}$, hence $\phi_\beta(0)=\frac{1}{2}\ln \beta+\phi_{\beta=1}(0)$.
Combining altogether we see that $\phi_\beta(x)=(\frac{1}{2}-\beta x)\ln{\beta}+c$, where
\begin{equation}\label{ZamoBarn2}
\fl c = \int_0^\infty \frac{dt}{t} \big( \frac{ e^{- t}}{1-e^{-t}}
+\frac{e^{-t}}{2}-\frac{1}{t} \big)\equiv \lim_{z\to 0}\int_0^\infty \frac{dt e^{-zt}}{t} \big( \frac{1}{e^{t}-1}
+\frac{1}{2}-\frac{1}{t} \big)+\int_0^\infty \frac{dte^{-zt}}{t}\frac{e^{-t}-1}{2}
\end{equation}
 which using \cite{GR} 8.341.1, p.888 yields
 \begin{equation}\label{ZamoBarn2}
\fl c =  \lim_{z\to 0}\left[\ln{\Gamma(z)}+z-(z-1/2)\ln{z}-\frac{1}{2}\ln{2\pi}+\int_0^\infty \frac{dte^{-zt}}{t}\frac{e^{-t}-1}{2}\right]
\end{equation}
The last integral is equal to
$\frac{1}{2}\ln{\left(z/(z+1)\right)}$, and after straightforwardly
taking the limit we find finally $c=-\frac{1}{2}\ln{2\pi}$, in full
agreement with (\ref{Gt1}).

Next we want to obtain the asymptotics. For this it is useful to note that:
\begin{eqnarray}\label{h1a}
h_\beta(x) := \partial_x^2 \ln G_\beta(x) = \int_0^\infty \frac{dt}{t} ( e^{-t} - \frac{t^2 e^{- x t}}{(1-e^{-\beta t})(1-e^{-t/\beta})} )
\end{eqnarray}
Exploiting again the identity (\ref{log}) we can rewrite the above formula in a form more convenient for applications, see Eq. (\ref{h1b}) in the text.
For example, by changing variables $t=\tau/x$ in (\ref{h1b}) we immediately can find the asymptotic behaviour for $x\to \infty$ at fixed $\beta$ to be given by
$$ h_\beta(x)=\ln{x}-\frac{1}{2x}\left(\beta+\frac{1}{\beta}\right)+\ldots $$ as long as $x\gg max(\beta,\beta^{-1})$.
The same asymptotic behaviour holds for $h_\beta(z)$ in the complex plane for  $\Re{z}>0$ and $|z|\to \infty$.

Note also the useful doubling formula (\ref{doubling}) for
$G_\beta(x)$ which we were not able to trace in the available
literature.

\section{High temperature expansions}

\subsection{high temperature expansion of REM models}

It is useful to derive high temperature expansions for a Gaussian REM
$Z_M(\beta) =\sum_{i=1}^M e^{- \beta V_i}$ with an arbitrary correlation matrix $\overline{V_i V_j}=C_{ij}$. One expands:
\begin{eqnarray}
&& Z_M(\beta) = M - \beta \sum_i V_i + \frac{1}{2} \beta^2 \sum_i V_i^2 -  \frac{1}{6} \beta^3 \sum_i V_i^3 + O(\beta^4)
\end{eqnarray}
which leads to:
\begin{eqnarray}
&& \ln Z_M(\beta) = \ln M - \frac{\beta}{M} \sum_i V_i + \frac{1}{2} \beta^2 ( \frac{1}{M} \sum_i V_i^2 - \frac{1}{M^2}
\sum_{ij} V_i V_j) \\
&& - \frac{1}{6} \beta^3 ( \frac{1}{M} \sum_i V_i^3 - 3 \frac{1}{M^2}
\sum_{ij} V_i V_j^2 + 2 \frac{1}{M^3} \sum_{ijk} V_i V_j V_k )+ O(\beta^4) \nonumber
\end{eqnarray}
This leads to the average free energy:

\begin{eqnarray}
&& F_M(\beta) = - \frac{1}{\beta} \overline{\ln Z_M(\beta)} = - \frac{1}{\beta} \ln M - \frac{1}{2} \beta (\frac{1}{M} \sum_i C_{ii} - \frac{1}{M^2}
\sum_{ij} C_{ij}) + O(\beta^3)
\end{eqnarray}
and the variance:
\begin{eqnarray}
&& \overline{f^2}^c = \frac{1}{\beta^2} \overline{ \ln^2 Z_M(\beta) - \overline{\ln Z_M(\beta)}^2}
=  \frac{1}{M^2} \sum_{ij} C_{ij} + \beta^2 \big( \frac{1}{M^2} \sum_{ij} (C_{ii} C_{ij} + \frac{1}{2} C_{ij}^2) \nonumber  \\
&& -  \frac{1}{M^3} \sum_{ijk} (3 C_{ij} C_{ik} + C_{ii} C_{jk} ) +
\frac{5}{2} \frac{1}{M^4} \sum_{ijkl} C_{ij} C_{kl} \big) +
O(\beta^4) \label{expand}
\end{eqnarray}
This result for $\overline{f^2}^c$ is useful to test universality in the sense (ii), i.e. at any temperature. Let us examine several cases.

Consider first the periodic case discussed in the text, where $C_{ij}$ is a cyclic (i.e. circulant) matrix, i.e. of the form (\ref{eigen}). Then $\frac{1}{M^2} \sum_{ij} C_{ij} = \lambda_0/M$. Fixing $\lambda_0=0$ as we did here, we find that
the expression for the second cumulant of the free energy simplifies and that it vanishes at $\beta=0$ as:
\begin{eqnarray}
&& \overline{f^2}^c = \frac{\beta^2}{2 M^2} \Tr C^2 + O(\beta^4) = \frac{\beta^2}{2 M^2} \sum_{k \neq 0} \lambda_k^2 + O(\beta^4)
\end{eqnarray}
It is now easy to check that both the discrete circular model (\ref{circmat}), the sharp model (SM) and the long range model (LRM) behave in the limit $M \to +\infty$ as:
\begin{eqnarray}
&& \overline{f^2}^c = \beta^2 \sum_{k=1}^\infty \frac{1}{k^2} + O(\beta^4) = \frac{\pi^2}{6} \beta^2 + O(\beta^4)
\end{eqnarray}
i.e. as the continuum circular model for which one has (\ref{f2circ2}). This is consistent with the conjecture that these models belong to the same universality class at any temperature. Furthermore the coefficient of $\beta^2$ can also be obtained, e.g. for the discrete circular ensemble (\ref{circmat}), as:
\begin{eqnarray}
&& \lim_{M \to \infty} \frac{1}{2 M^2} \sum_{ij} C_{ij}^2 = \frac{1}{2} \int_0^{2 \pi} \frac{d\theta_1}{2 \pi} \int_0^{2 \pi} \frac{d\theta_2}{2 \pi} [ 2 \ln 2 |\sin( \frac{\theta_1-\theta_2}{2} )| ]^2 = \frac{\pi^2}{6}
\end{eqnarray}
Note that the diagonal does not contribute to this limit (its contribution is $O(\ln^2 M/M)$ and that will be a general fact.

For the discrete interval model (\ref{intervalM}) we compute the two first terms in the high temperature expansion. We see that the terms involving $C_{ii}$ cancel out, as it should. In the limit $M\to \infty$ we replace remaining sums by integrals and get
\begin{eqnarray}
&& \overline{f^2}^c =-2I_0+2\beta^2\left[I_1+5I_0^2-6\tilde{I}\right]+O(\beta^4)\,,
\end{eqnarray}
where we have defined the integrals:
\begin{eqnarray} \label{beta0direct}
 I_0 &=&\int_0^{1} dx_1 \int_0^{1} dx_2 \ln|x_2-x_1|, \quad I_1=\int_0^{1} dx_1 \int_0^{1} dx_2 \ln^2|x_2-x_1| \\
&& \tilde{I}=\int_0^{1} dx_1 \int_0^{1} dx_2 \int_0^{1} dx_3 \ln|x_2-x_1|  \ln|x_3-x_1|
\end{eqnarray}
Calculation of these integrals give:
\[
I_0 = -3/2 \quad , \quad I_1=\frac{7}{2},\quad \tilde{I}=\frac{17}{6}-\frac{\pi^2}{18}
\]
which gives the final result:
\begin{eqnarray}\label{expancorr}
&& \overline{f^2}^c =3+\beta^2\left[\frac{2}{3}\pi^2-\frac{9}{2}\right]+O(\beta^4)\,,
\end{eqnarray}
As we show below this coincides with our analytical prediction from the continuum model, see (\ref{f2n}) below.

As we see in Fig. \ref{fig:f2} for the discrete interval model (\ref{intervalM}) at finite $M$
$\overline{f^2}^c(\beta=0)$ is smaller than $3$. In fact, one can add a $W_M$ on the diagonal in (\ref{intervalM}) so as to tune this value to exactly $3$ for any $M$, without changing the universality class (i.e. $W_M$ goes to zero fast enough). One could try to systematize this idea, e.g. to add to the correlation matrix of the discrete model some other matrix, subdominant in the limit $M \to \infty$, so as to fit the lowest orders coefficients in $\beta^p$ to their actual value for the continuum model - those are given below for the interval, see formula (\ref{f2n}). We have checked for the circular case that it can be easily implemented up to $p=2$. Whether this will allow to select better discrete models with faster convergence even at lower temperature is left for future studies.

Concerning the class of model (\ref{1}), we can similarly check that for the associated discrete REM, i.e. $C_{ij} = - 2 \ln|2 \sin(\frac{1}{2} f(\theta_i)-\frac{1}{2} f(\theta_j))|$ for $i \neq j$ and $\theta_i=2 \pi i/M$ one has:
\begin{eqnarray}
&& \! \! \! \! \! \! \! \! \! \! \! \! \! \! \! \!  \lim_{M \to \infty} \frac{1}{M^2} \sum_{ij} C_{ij} = - 2 \int_0^{2 \pi} \frac{d\theta_1}{2 \pi} \int_0^{2 \pi} \frac{d\theta_2}{2 \pi} \ln [ 2 |\sin(\frac{f(\theta_1)-f(\theta_2)}{2})| ] = \frac{1}{2} a^2 + O(a^4)
\end{eqnarray}
for $f(\theta)=\theta+a \sin(\theta)$, hence at small $a$, $\overline{f^2}^c=\frac{1}{2} a^2 + O(a^4) + O(\beta^2)$
as announced in the text, and there is no universality valid at all temperature (the universality class in the sense (ii) defined above depends on the function $f(\theta)$).

Finally the same expansion (\ref{expand}) holds for any continuum REM of the form $Z= \int dx \rho(x) e^{- \beta V(x)}$ and can be obtained from the above just replacing $\frac{1}{M^n} \sum_{i_1,..i_n} \to \frac{1}{(\int dx \rho(x))^n} \int_{x_1,..x_n}$ and $C_{i_1,i_2}$ by its continuum expression $C(x_1,x_2)$.

\subsection{high temperature expansion of the analytical result for the interval}

Let us derive the high temperature expansion of our analytical result (\ref{f2int})-(\ref{h1b}) for the second cumulant of the free energy. Since an independent method also exists to obtain this expansion, as displayed in the discussion above, this constitutes a check of our solution in the high temperature phase. For this we need to use (\ref{h1b}) for $x=\alpha_1 \beta +\alpha_2\frac{1}{\beta}$, where $\alpha_{1,2}$ are given positive constants. By introducing $\tau=t/\beta$ we have:
\begin{eqnarray}
\fl h_\beta\left(\alpha_1\,\beta+\frac{\alpha_2}{\beta}\right)=\ln{\frac{\alpha_2}{\beta}}+\ln{\left(1+\frac{\alpha_1}{\alpha_2}\,\beta^2\right)}+
\int_0^\infty \frac{d\tau}{\tau}  e^{-\alpha_2\tau-\alpha_1\beta^2\tau} \left(1-\frac{\tau^2\beta^2}{(1-e^{-\beta^2 \tau})(1-e^{-\tau})} \right) \nonumber  \\
&&
\end{eqnarray}
which can be easily used to expand in powers of $\beta^2$. In particular, the leading term from  (\ref{f2int}) is a constant given by:
\begin{eqnarray}
\fl \overline{f^2}^c = 2\ln{3}+\ln{2} + \int_0^{\infty}\frac{d\tau}{\tau}\left(1-\frac{\tau}{1-e^{-\tau}} \right)\left(
2e^{-\frac{3}{2}\tau}-e^{-\tau}-e^{-2\tau}\right)
\end{eqnarray}
where the integral can be computed by pieces using the formulas (see 3.311.7 \cite{GR}):
\begin{equation}
\int_{0}^{\infty}d\tau\,\frac{e^{-\mu\tau}-e^{-\nu\tau}}{1-e^{-\tau}}=\psi(\nu)-\psi(\mu)
,\quad \mbox{and}\quad
\int_0^\infty \frac{d\tau}{\tau}\left(e^{-\mu\tau}-e^{-\nu\tau}\right)=\ln{\frac{\nu}{\mu}}
\end{equation}
combining altogether we find:
\begin{eqnarray}  \label{beta0analytic}
\overline{f^2}^c(\beta=0) = 4\ln{2}-[\psi(1)+\psi(2)-2\psi(3/2)]=3
\end{eqnarray}
in agreement with the result obtained above in (\ref{beta0direct}) by a direct method.

This expansion can be carried to higher order. Using mathematica and some heuristics we find that it can
be put in the form:
\begin{eqnarray} \label{f2n}
&& \overline{f^2}^c = 3 + ( \frac{2}{3} \pi^2 - \frac{9}{2}) \beta^2 + \sum_{k=2}^\infty
 (-1)^{k+1} 3 k ~(\zeta(k+1)- (1 + \frac{B_k}{k}) ) ~\beta^{2 k}
\end{eqnarray}
where the $B_k$ are the Bernouilli numbers ($B_k=0$ for $k$ odd).



\begin{thebibliography}{99}
\bibitem{footnote} equivalently one can choose $\epsilon=1/M$ and $W=-2 \ln R$ which, as we know, does not change the distribution of the maximum
\end{thebibliography}
\end{document}